\newcommand*{\COMMENTSOFF}{} 
\newcommand*{\PUBLICATION}{} 

\documentclass[conference]{IEEEtran}

\newcommand{\attack}{JackHammer} 
\newcommand{\para}[1]{\smallskip\noindent\textbf{{#1.}}}

\author{
  \IEEEauthorblockN{
    Zane Weissman\IEEEauthorrefmark{1},  
    Thore Tiemann\IEEEauthorrefmark{2},
    Daniel Moghimi\IEEEauthorrefmark{1},
    Evan Custodio\IEEEauthorrefmark{3},
    Thomas Eisenbarth\IEEEauthorrefmark{2}\IEEEauthorrefmark{1} and
    Berk Sunar\IEEEauthorrefmark{1}
}

\IEEEauthorblockA{\IEEEauthorrefmark{1}Worcester Polytechnic Institute,
Worcester, MA, USA}

\IEEEauthorblockA{\IEEEauthorrefmark{2}University of L\"ubeck, L\"ubeck, Germany}

\IEEEauthorblockA{\IEEEauthorrefmark{3}Intel Corporation, Hudson, MA, USA}
}

\makeatletter

\usepackage{comment}
\usepackage{breakcites}
\usepackage{makecell}
\usepackage{listings}
\usepackage{color}
\usepackage{algorithmicx}
\usepackage{tabularx}
\usepackage[ruled]{algorithm}
\usepackage[noend]{algpseudocode}
\usepackage{algpseudocode} 
\usepackage{amsmath}
\usepackage{graphicx}
\usepackage{subcaption}
\usepackage{amssymb}
\usepackage{epstopdf}
\usepackage{tikz} 
\usepackage{enumitem}
\usetikzlibrary{arrows, automata, calc, positioning}
\usepackage{pgfplots}
\usepackage{environ}
\usepackage{units}
\usepackage[colorinlistoftodos,prependcaption,textsize=tiny]{todonotes}
\PassOptionsToPackage{hyphens}{url}\usepackage{hyperref}
\usepackage{multirow}
  \definecolor{linkcolor}{rgb}{0.25,0.1,0.7}
  \definecolor{citecolor}{rgb}{0,0.4,0}
  \definecolor{urlcolor}{rgb}{0,0,0.65}
  \hypersetup{pdfpagemode=UseNone,pdfstartview=FitH,colorlinks=true, linkcolor=linkcolor, urlcolor=urlcolor, citecolor=citecolor}
\usepackage{xcolor}
\usepackage{colortbl}

\usepackage{booktabs}

\newenvironment{CompactItemize}%
{\begin{list}{$\blacktriangleright$}%
  {\leftmargin=\parindent \itemsep=2pt \topsep=2pt
    \parsep=0pt \partopsep=0pt}}%
{\end{list}}

\definecolor{dkgreen}{rgb}{0,0.6,0}
\definecolor{gray}{rgb}{0.5,0.5,0.5}
\definecolor{mauve}{rgb}{0.58,0,0.82}
\ifdefined\COMMENTSOFF
  \newcommand\daniel[1]{}
  \newcommand\berk[1]{}
  \newcommand\zane[1]{}
  \newcommand\thore[1]{}
\else
  \newcommand\daniel[1]{\todo[color=green]{DanM: #1}}
  \newcommand\berk[1]{\todo[color=purple]{Berk: #1}}
  \newcommand\zane[1]{\todo[color=red]{Zane: #1}}
  \newcommand\thore[1]{\todo[color=yellow]{Thore: #1}}
\fi

\begin{document}
\title{
\ifdefined\SUBMISSION
{Resubmission:} 
\fi
\attack:
Efficient Rowhammer on Heterogeneous FPGA-CPU Platforms
}
\maketitle
\thispagestyle{plain}
\pagestyle{plain}

\begin{abstract}
After years of development, FPGAs are finally making an appearance on multi-tenant cloud servers. 
Heterogeneous FPGA-CPU microarchitectures require reassessment of common assumptions about isolation and security boundaries,
as they introduce new attack vectors and vulnerabilities. 
In this work, we analyze the memory and cache subsystem and study Rowhammer and cache attacks enabled by two proposed heterogeneous FPGA-CPU platforms from Intel: 
the \textsf{Arria 10 GX} with an \textsf{integrated} FPGA-CPU platform, 
and the \textsf{Arria 10 GX} PAC expansion card which connects the FPGA to the CPU via the PCIe interface. 
We demonstrate \attack, a novel, efficient, and stealthy Rowhammer from the FPGA to the host's main memory. 
Our results indicate that a malicious FPGA can perform {\bf twice as fast} as a typical Rowhammer from the CPU on the same system and causes around {\bf four times as many} bit flips as the CPU attack.
We demonstrate the efficacy of \attack\ from the FPGA through a realistic fault attack on the \textsf{WolfSSL} RSA signing implementation that reliably causes a fault after an average of {\bf fifty-eight RSA signatures}, {\bf 25\% faster} than a CPU Rowhammer.
In some scenarios our \attack\ attack produces faulty signatures {\bf more than three times more often} and {\bf almost three times faster} than a conventional CPU Rowhammer.
Finally, we systematically analyze new cache attacks in these environments following demonstration of a cache covert channel across FPGA and CPU.
\end{abstract}

  
\section{Introduction}
In recent years, as improvements in the performance of microprocessors have slowed, developers have looked to other computing resources. 
FPGAs, graphics processing units (GPUs) and application-specific integrated circuits (ASICs) have all been adapted to accelerate applications such as high-frequency trading and machine learning. 
FPGAs are particularly interesting for cloud computing applications, as they can be reconfigured for the needs of different users at different times.
Amazon Web Services~\cite{amazon2019ec2f1} and Alibaba Cloud~\cite{alibaba2019fpgacloud} already offer FPGA instances with ultra-high performance Xilinx Virtex UltraScale+ and \textsf{Intel Arria 10 GX} FPGAs to the consumer market. 
These FPGAs are designed for high I/O bandwidth and high compute capacity, making them ideal for server workloads. 
New Intel FPGAs offer cache-coherent memory systems for even better performance when data is being passed back and forth between CPU and FPGA. 

The flexibility of FPGA systems can also open up new attack vectors for malicious users in public clouds or more efficiently exploit existing ones. 
\emph{Integrated FPGA platforms} connect the FPGA right into the processor bus interconnect giving the FPGA direct access into cache and memory~\cite{intel2018integrated}.
Similarly, high-end FPGAs can be integrated into a server as an accelerator, e.g.\ connected via PCIe interface~\cite{intel2018pac,xilinx2019accelerator}.
Such combinations provide unprecedented performance over a high-throughput and low-latency connection with the versatility of a reprogrammable FPGA infrastructure shared among cloud users. 
However, the tight integration may also expose users to new adversarial threats.

This work exposes hardware vulnerabilities in hybrid FPGA-CPU systems with a particular focus on cloud platforms where the FPGA and the CPU are in distinct security domains: one potentially a victim and the other an attacker. 
We examine Intel's \textsf{Arria 10 GX} FPGA as an example of a current generation of FPGA accelerator platform designed in particular for heavy and/or cloud-based computation loads. 
We thoroughly analyze the memory interfaces between such platforms and their host CPUs. 
These interfaces, which allow the CPU and FPGA to interact in various direct and indirect ways, include 
hardware on both the FPGA and CPU, application libraries and software drivers executed by the CPU, and logical interfaces implemented on the FPGA outside of but accessible to the user-configurable region.
We propose attacks that exploit practical use cases of these interfaces to target adjacent systems such as the CPU memory and cache.
  
\subsection{Our Contributions}
We demonstrate novel attacks between the memory interface of \textsf{Intel Arria 10 GX} platforms and their host CPUs. 
Furthermore, we demonstrate a Rowhammer mounted from the FPGA against the CPU to cause faults in the \textsf{WolfSSL} RSA signature implementation, and to leak a private RSA modulus factor. 
In summary:

\begin{CompactItemize}
\item We thoroughly reverse-engineer and analyze the cache behavior and investigate the viability of cache attacks on realistic FPGA-CPU hybrid systems. 

\item Based on our study of the cache subsystem, we build \attack, a Rowhammer from the FPGA that bypasses caching to hammer the main memory.
We compare \attack\ with the CPU Rowhammer and show that \attack\ is twice as fast as a CPU attack, 
causing faults that the CPU Rowhammer is unable to replicate. \attack\ remains stealthy to CPU monitors since it bypasses the CPU microarchitecture.

\item Using both \attack\ and conventional CPU Rowhammer, we demonstrate a fault attack on recent versions of RSA implementation in the \textsf{WolfSSL} library and recover private keys. 
We show that the base blinding used in this RSA implementation leaves the algorithm vulnerable to the Bellcore fault injection attack. 

\item We systematically analyze cache attack techniques on different scenarios: FPGA to CPU, CPU to FPGA, and FPGA to FPGA,
 and demonstrate a cache covert channel that can transmit up to \unit[1.5 -- 1.8]{MBit/s} from the FPGA to the CPU.

\end{CompactItemize}

\para{Vulnerability Disclosure}
We informed the \textsf{WolfSSL} team about the vulnerability to Bellcore-style RSA fault injection attacks on November 25, 2019. 
\textsf{WolfSSL} acknowledged the vulnerability on the same day, and released \textsf{WolfSSL} 4.3.0 with a fix for the vulnerability on December 20, 2019. 
The vulnerability can be tracked via CVE-2019-19962~\cite{mitre2019wolfssl}.

\subsection{Experimental Setup}

We experiment with two distinct FPGA-CPU platforms with \textsf{Intel Arria 10} FPGAs: 
\textbf{1)} integrated into the CPU package and 
\textbf{2)} Programmable Acceleration Card (PAC).
The integrated \textsf{Intel Arria 10} is based on a prototype E5-2600v4 CPU with 12 physical cores.
The CPU has a Broadwell architecture in which the last level cache (LLC) is inclusive of the L1/L2 caches.
The CPU package has an integrated \textsf{Arria 10 GX 1150} FPGA running at \unit[400]{MHz}.
All measurements done on this platform are strictly done from userspace only,
as access is kindly provided by Intel through their Intel Lab (IL) Academic Compute Environment.\footnote{\url{https://wiki.intel-research.net/}}
The IL environment also gives us access to platforms with PACs with \textsf{Arria 10 GX 1150} FPGA installed and running at \unit[200]{MHz}. 
These systems have Intel Xeon Platinum 8180 CPUs that come with non-inclusive LLCs. 
We carried out the Rowhammer experiments on our local Dell Optiplex 7010 system with an Intel i7-3770 CPU, 
and a single DIMM of Samsung M378B5773DH0-CH9 \unit[1333]{MHz} \unit[2]{GB} DDR3 DRAM equipped with the same Intel PAC running
with a primary clock speed of \unit[200]{MHz}.\footnote{The PAC is intended to support \unit[400]{MHz} clock speed, 
but the current version of the Intel Acceleration Stack has a bug that halves the clock speed.}

The operating system (OS) running in the IL is a 64-bit Red Hat Enterprise Linux 7 with Kernel version 3.10.
The \textsf{OPAE} version was compiled and installed on July 15th, 2019 for both the FPGA PAC and the integrated FPGA platform. 
We used Quartus 17.1.1 and Quartus 16.0.0 to synthesize AFUs for the PACs and integrated FPGAs, respectively. 
The bitstream version of the non-user-configurable Board Management Controller (BMC) firmware is 1.1.3 on the FPGA PAC and 5.0.3 on the integrated FPGA.
The OS on our Optiplex 7010 workstation is Ubuntu 16.04.4 LTS with Linux kernel 4.13.0-36. 
On this system, we installed the latest stable release of \textsf{OPAE}, 1.3.0, and on its FPGA PAC, we installed the compatible 1.1.3 BMC firmware bitstream.

\section{Background}

\subsection{RSA-CRT Signing}
RSA signatures are computed by raising a plaintext $m$ to a secret power $d$ modulo $N = pq$, where $p$ and $q$ are prime and secret, and $N$ is public~\cite{boneh1997importance}. 
These numbers must all be large for RSA to be secure, which makes the exponentiation rather slow.
However, there is an algebraic shortcut for modular exponentiation: 
the Chinese Remainder Theorem (CRT), used in many RSA implementations, including in the \textsf{WolfSSL} we attack in \autoref{sec:rsa_fault_Rowhammer} and in OpenSSL~\cite{carre2018openssl}. 
The basic form of the RSA-CRT signature algorithm is shown in \autoref{alg:rsa-crt}.
The CRT algorithm is much faster than simply computing $m^d \bmod N$ because $d_p$ and $d_q$ are of order $p$ and $q$ respectively while $d$ is of order $N$, which, being the product of $p$ and $q$, is significantly greater than $p$ or $q$; 
it is around four times faster to compute the two exponentiations $m^{d_p}$ and $m^{d_q}$ than it is to compute $m^d$ outright~\cite{aumuller2002fault}. 

\begin{algorithm}[tb]
  \caption{Chinese remainder theorem RSA signature.}
  \label{alg:rsa-crt}
  \begin{algorithmic}[1]
  \Procedure{sign}{$m$: message, $d$: private exponent, $p$: private factor, $q$: private factor}
  \State $S_p \gets m^{d_p}\bmod p$\Comment{equivalent to $m^d \bmod p$}
  \State $S_q \gets m^{d_q}\bmod q$\Comment{equivalent to $m^d \bmod q$}
  \State $I_q \gets q^{-1}\bmod p$ \Comment{inverse of $q$}
  \State \textbf{return} $S \gets S_q + q \left((S_p-S_q)I_q\bmod p\right)$
  \EndProcedure
  \end{algorithmic}
\end{algorithm}

\subsection{Cache Attacks}
Cache attacks have been proposed attacking various applications
~\cite{oren2015spy,brumley2009cache,gulmezoglu2017cache,gulmezoglu2017perfweb,benger2014ooh,tsunoo2003cryptanalysis}. 
In general, cache attacks use timing of the cache behavior to leak information. 
Modern cache systems use a hierarchical architecture that includes smaller, faster caches and bigger, slower caches. 
Measuring the latency of a memory access can often confidently determine which levels of cache contain a certain memory address (or if the memory is cached at all).
Many modern cache subsystems also support coherency, which ensures that whenever memory is overwritten in one cache, 
copies of that memory in other caches are either updated or invalidated.
Cache coherency may allow an attacker to learn about a cache line that is not even directly accessible~\cite{irazoqui2016cross}.
Cache attacks have become a major focus of security research in cloud computing platforms where users are allocated CPUs, 
cores, or virtual machines which, in theory, should offer perfect isolation, but in practice may leak information to each other via shared caches~\cite{inci2016cache}.
An introduction to various cache attack techniques is given below:

\para{Flush+Reload}
Flush+Reload (F+R)~\cite{yarom2014flush} has three steps: 
\textbf{1)} The attacker uses the \texttt{clflush} instruction to flush the cache line that is to be monitored.
After flushing this cache line, \textbf{2)} she waits for the victim to execute.
Later, \textbf{3)} she reloads the flushed line and measures the reload latency.
If the latency is low, the cache line was served from the cache hierarchy, so the cache line was accessed by the victim.
If the access latency is high, the cache line was loaded from main memory, meaning that the victim did not access it.
F+R can work across cores and even across sockets, as long as the LLC is coherent, as is the case with many modern multi-CPU systems.
Flush+Flush (F+F)~\cite{gruss2016flush} is similar to F+R, 
but the third step is different: the attacker flushes the cache line again and measures the execution time of the flush instruction instead of the memory access.

Orthogonal to F+R, if the attacker does not have access to an instruction to flush a cache line, 
she can instead evict the desired cache line by accessing cache lines that form an eviction set
in an Evict+Reload (E+R)~\cite{lipp2016armageddon} attack. 
Eviction sets are described shortly.
E+R can be used if the attacker shares the same CPU socket (but not necessarily the same core) as the victim and if the LLC is inclusive.\footnote{
A lower-level cache is called inclusive of a higher-level cache if all cache lines present in the higher-level cache are always present in the lower-level cache.} 
F+R, F+F, and E+R are limited to shared memory scenarios, where the victim and attacker share data or instructions, e.g.\ when memory de-duplication is enabled.

\para{Prime+Probe}
Prime+Probe (P+P) gives the attacker less temporal resolution than the aforementioned methods since the attacker checks the status of the cache by probing a whole cache set rather than flushing or reloading a single line. 
However, this resolution is sufficient in many cases~\cite{osvik2006cache,ristenpart2009hey,zhang2012cross,irazoqui2015s,oren2015spy,lipp2016armageddon,moghimi2017cachezoom}.
P+P has three steps: 
\textbf{1)} The attacker primes the cache set under surveillance with dummy data by accessing a proper eviction set,
\textbf{2)} she waits for the victim to execute,
\textbf{3)} she accesses the eviction set again and measures the access latency (probing).
If the latency is above a certain threshold, some parts of the eviction set was evicted by the victim process, 
meaning that the victim accessed cache lines belonging to the cache set under surveillance~\cite{liu2015last}.
Unlike F+R, E+R, and F+F, P+P does not rely on shared memory.
However, it is noisier, only works if the victim is located on the same socket as the attacker, and relies on inclusive caches.
An alternative attack against non-inclusive caches is to target the cache directory structure~\cite{yan2019directories}.

\para{Evict+Time} 
In scenarios where the attacker can not probe the target cache set or line, but she can still influence the target cache line, 
an Evict+Time (E+T) is still possible depending on the target application. 
In an E+T attack, the attacker only evicts the victim's cache line and measures the aggregate execution time of the victim's operation, 
hoping to observe a correlation between the execution time of an operation such as a cryptographic routine and the cache access pattern.  

\para{Eviction Sets}
Caches store data in units of cache lines that can hold $2^b$ bytes each (64 bytes on Intel CPUs).
Caches are divided into $2^s$ sets, each capable of holding $w$ cache lines.
$w$ is called the way-ness or associativity of the cache.
An eviction set is a set of congruent cache line addresses capable of filling a whole cache set.
Two cache lines are considered congruent if they belong to the same cache set.
Memory addresses are mapped to cache sets depending on the $s$ bits of the physical memory address directly following the $b$ cache line offset bits, which are the least significant bits.
Additionally, some caches are divided into $n$ slices, where $n$ is the number of CPU cores.
In the presence of slices, each slice has $2^s$ sets with $w$ ways each.
Previous work has reverse-engineered the mapping of physical address bits to cache slices on some Intel processors~\cite{irazoqui2015systematic}. 
A minimal eviction set contains $w$ addresses and therefore fills an entire cache set when accessed. 

\subsection{Rowhammer}
DRAM cells discharge over time, and the memory controller has to refresh the cells to avoid accidental data corruption. 
Generally, DRAM cells are laid out in banks and rows, and each row within a bank has two adjacent rows, one on either side. 
In a Rowhammer attack, memory addresses in the same bank as the target memory address are accessed in quick succession. 
When memory adjacent to the target is accessed repeatedly, the electrostatic interference generated by the physical process of accessing the memory can elevate the discharge for bits stored in the target memory. 
A ``single-sided'' Rowhammer performs accesses to just one of these rows to generate bit flips in the target row; 
a ``double-sided'' Rowhammer performs accesses to both adjacent rows and is generally more effective in producing bit flips.
Rowhammer relies on the ability to find blocks of memory accessible to the malicious program (or in this work, hardware) that are in the same memory bank as a given target address. 
The standard way of finding these memory addresses is by exploiting row buffer conflicts as a timing side-channel~\cite{frigo2018grand}.
Pessl et al.~\cite{pessl2016drama} reverse-engineered the bank mapping algorithms of several CPU and DRAM configurations which allows an attacker to deterministically calculate all of the physical addresses that share the same bank if the chipset and memory configuration are known.

\subsection{Related Attacks}
Classical power analysis techniques like Kocher et al.'s differential power analysis~\cite{kocher1999differential} have been applied in new attacks on inter-chip FPGAs~\cite{zhao2018fpga,schellenberg2018remote,schellenberg2018inside}. 
Such integrated and inter-chip FPGAs are available in various cloud environments and system-on-chips (SoCs) products. 
In particular, Zhao et al.~\cite{zhao2018fpga} demonstrated how to build an on-chip power monitor using ring oscillators (ROs) which can be used to attack the host CPU or other FPGA tenants.
In multi-tenant FPGA scenarios where partial reconfiguration by two separate security domains is possible, more powerful attacks become possible. 
For instance, the long wires on the FPGA can spy on adjacent wires using ROs~\cite{giechaskiel2019leakier,ramesh2018fpga,provelengios2019characterization}. 
Ramesh et al.~\cite{ramesh2018fpga} exploited the speed of ROs to infer the carried bit in the adjacent wire and demonstrated a key recovery attack on AES. 
ROs can also be used as power wasters to create voltage drop and timing faults~\cite{gnad2017voltage,krautter2018fpgahammer}.
Note that such attacks rely on FPGA multi-tenancy which is not widely used yet.
In contrast, in this work, we only assume that the FPGA-CPU memory subsystem is shared among tenants.

\section{Analysis of Intel FPGA-CPU Systems}
\begin{figure}[tp]
  \centering
  \scalebox{.8}{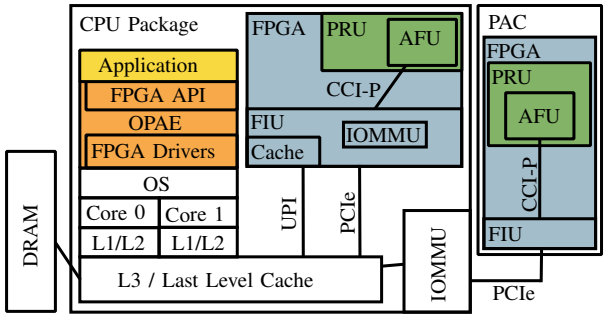}
  \caption{
  Overview of the architecture of Intel FPGAs. 
  The software part of the Intel Acceleration Stack called \textsf{\textsf{OPAE}} is highlighted in orange. 
  Its API is used by applications (yellow) to communicate with the AFU. 
  The Green Region marks the part of the FPGA that is re-configurable from userspace at runtime. 
  The Blue Region describes the static soft core of the FPGA.
  It exposes the \textsf{CCI-P} interface to the AFU.}
  \label{fig:attackscenario} 
\end{figure}

This section explains the hardware and software interfaces that the \textsf{Intel Arria 10 GX} FPGA platforms use to communicate with their host CPUs and the firmware, drivers, and architectures that underlay them.
\autoref{fig:attackscenario} gives an overview of such architecture.

\para{Introduction to Intel Terminology}
Intel refers to a single logical unit implemented in FPGA logic and having a single interface to the CPU as an Accelerator Functional Unit (AFU). 
So far, available FPGA platforms only support one AFU per Partial Reconfiguration Unit (PRU, also called the \emph{Green Region}).
The AFU is an abstraction similar to a program that captures the logic implemented on an FPGA. 
The FPGA Interface Manager (FIM) is part of the non-user-configurable portion (\emph{Blue Region}) of the FPGA and contains external interfaces 
like memory and network controllers as well as the FPGA Interface Unit (FIU), 
which bridges those external interfaces with internal interfaces to the AFU.

\subsection{Intel FPGA Platforms}

\textbf{\textsf{Intel's Arria 10 GX} Programmable Acceleration Card (PAC)} is a PCIe expansion card for FPGA acceleration~\cite{intel2018pac}. 
The \textsf{Arria 10 GX} FPGA on the card communicates with its host processor over a single PCIe Gen3x8 bus. 
Memory reads and writes from the FPGA to the CPU's main memory use physical addresses; 
in virtual environments, the PCIe controller on the CPU side implements an I/O memory management unit (IOMMU) to translate physical addresses 
in the virtual machine (what Intel calls I/O Virtual Addresses or IOVA) to physical addresses in the host.
Alongside the FPGA, the PAC carries \unit[8]{GB} of DDR4, \unit[128]{MB} of flash memory, and USB for debugging.

An alternative accelerator platform is the \textbf{Xeon server processor with an integrated \textsf{Arria 10} FPGA} in the same package~\cite{intel2018integrated}.
The FPGA and CPU are closely connected through two PCIe Gen3x8 links and an UltraPath Interconnect (UPI) link. 
UPI is Intel's high-speed CPU interconnect (replacing its predecessor QPI) in Skylake and later Intel CPU architectures~\cite{mulnix2017skylake}.
The FPGA has a \unit[128]{KiB} directly mapped cache that is coherent with the CPU caches over the UPI bus. 
Like the PCIe link on the PAC, both the PCIe links and the UPI link use I/O virtual addressing, appearing as physical addresses to virtualized environments. 
As the UPI link bypasses the PCIe controller's IOMMU, the FIU implements its own IOMMU and Device TLB to translate physical addresses for reads and writes using UPI~\cite{intel2018ccip}.

\subsection{Intel's FPGA-CPU Compatibility Layers} 

\para{Open Programmable Acceleration Engine (\textsf{OPAE})}
Intel's latest generations of FPGA products are designed for use with the \textsf{OPAE}~\cite{intel2017opae} 
which is part of the Intel Acceleration Stack.
The principle behind \textsf{OPAE} is that it is an open-source, hardware-flexible software stack designed 
for interfacing with FPGAs that use Intel's Core Cache Interface (\textsf{CCI-P}), 
a hardware host interface for AFUs that specifies transaction requests, header formats, timing, and memory models~\cite{intel2018ccip}.
\textsf{OPAE} provides a software interface for software developers to interact with a hosted FPGA, while \textsf{CCI-P} provides a hardware interface for hardware developers to interact with a host CPU.
Excluding a few platform-specific hardware features, any \textsf{CCI-P} compatible AFU should be synthesizable (and the result should be logically identical) for any \textsf{CCI-P} compatible FPGA platform; 
\textsf{OPAE} is built on top of hardware- and OS-specific drivers and as such is compatible with any system with the appropriate drivers available. 
As described below, the \textsf{OPAE}/\textsf{CCI-P} system provides two main methods for passing data between the host CPU and the FPGA. 

\para{Memory-mapped I/O (MMIO)}
\textsf{OPAE} can send 32- or 64-bit MMIO requests to the AFU directly 
or it can map an AFU's MMIO space to OS virtual memory
~\cite{intel2017opae}. 
\textsf{CCI-P} provides an interface for incoming MMIO requests and outgoing MMIO read responses. 
The AFU may respond to read and write requests in any way that the developer desires, though an MMIO read request will time out after 65,536 cycles of the primary FPGA clock. 
In software, MMIO offsets are counted as the number of bytes and expected to be multiples of 4 (or 8, for 64-bit reads and writes), but in \textsf{CCI-P}, 
the last two bits of the address are truncated, because at least 4 bytes are always being read or written. 
There are 16 available address bits in \textsf{CCI-P}, meaning that the total available MMIO space is $2^{16}$ 32-bit words, or \unit[256]{KiB}~\cite{intel2018ccip}.

\para{Direct memory access (DMA)}\label{sec:dma}
\textsf{OPAE} can 
request the OS to allocate a block of memory that can be read by the FPGA. 
There are a few important details in the way this memory is allocated: most critically, it is allocated in a contiguous physical address space. 
The FPGA will use physical addresses to index the shared memory, so physical and virtual offsets within the shared memory must match. 
On systems using Intel Virtualization Technology for Directed I/O (VT-d), which employs the IOMMU to provide an IOVA to PCIe devices, 
the memory will be allocated in continuous IOVA space. 
Either way, this ensures that the FPGA will see an accessible and continuous buffer of the requested size. 
For buffer sizes up to and including one \unit[4]{kB} memory page, a normal memory page will be allocated to the calling process by the OS and configured to be accessible by the FPGA with its IOVA or physical address. 
For buffer sizes greater than \unit[4]{kB}, \textsf{OPAE} will call the OS to allocate a \unit[2]{MB} or \unit[1]{GB} huge page. 
Keeping the buffer in a single page ensures that it will be contiguously allocated in physical memory.

\subsection{Cache and Memory Architecture on the Intel FPGAs}
\label{sec:background_mem_cache_architecture}
\para{Arria 10 PAC}
The \textsf{Arria 10} PAC has  access to the CPU's memory system as well as its own local DRAM 
with a separate address space from that of the CPU and its memory. 
The PAC's local DRAM is always directly accessed, without a separate caching system.
When the PAC reads from the CPU's memory, the CPU's memory system will serve the request from its LLC if possible.
If the memory that is read or written is not present in the LLC, the request will be served by the CPU's main DRAM.
The PAC is unable to place cache lines into the LLC with reads, but writes from the PAC update the LLC.
\daniel{According to this, evict+time attack from the PAC to CPU should also be possible.}

\para{Integrated Arria 10}
The integrated \textsf{Arria 10} FPGA has access to the host memory. 
Additionally, it has its own \unit[128]{kB} cache that is kept coherent with the CPU's caches over UPI. 
Memory requests over PCIe take the same path as requests issued by an FPGA PAC. 
If the request is routed over UPI, the local coherent FPGA cache is checked first, on a cache miss, 
forwarding the request to the CPU's LLC or main memory.

\begin{table}[htb!]
	\centering
  \caption{Overview of the caching hints configurable over \textsf{CCI-P} on an integrated FPGA. 
  \texttt{*\_I} hints invalidate a cache line in the local cache. 
  Reading with \texttt{RdLine\_S} stores the cache line in the shared state. 
  Writing with \texttt{WrLine\_M} caches the line modified state.}
  \label{tab:cachinghints}
  \begin{tabularx}
    {\linewidth}
    {@{}p{.145\linewidth}| X X X X X}
    \toprule
		Cache Hint & \texttt{RdLine\_I} & \texttt{RdLine\_S} & \texttt{WrLine\_I} & \texttt{WrLine\_M} & \texttt{WrPush\_I} \\
		\midrule
		Desc.
		& \raggedright No FPGA caching
		& \raggedright Leave FPGA cache in S state
		& \raggedright No FPGA caching
		& \raggedright Leave FPGA cache in M state
		& \raggedright\arraybackslash Intent to cache in LLC \\\hline    
    Available & UPI, PCIe & UPI & UPI, PCIe & UPI & UPI, PCIe  \\
    \bottomrule     
  \end{tabularx}
\end{table}

\subsubsection{Reverse-engineering Caching Hint Behavior}\label{sec:caching_hints}
An AFU on the \textsf{Arria 10 GX} can exercise some control over caching behavior by adding caching hints to memory requests.
The available hints are summarized in \autoref{tab:cachinghints}.
For memory reads, \texttt{RdLine\_I} is used to not cache data locally and 
\texttt{RdLine\_S} to cache data locally in the shared state. 
For memory writes, \texttt{WrLine\_I} is used to prevent local caching on the FPGA, 
\texttt{WrLine\_M} leaves written data in the local cache in the modified state.  
\texttt{WrPush\_I} does not cache data locally but hints the cache controller to cache data in the CPU's LLC. 
The \textsf{CCI-P} documentation lists all caching hints as available for memory requests over UPI~\cite{intel2018ccip}. 
When sending requests over PCI, only \texttt{RdLine\_I}, \texttt{WrLine\_I}, and \texttt{WrPush\_I} can be used while other hints are ignored. 
However, based on our experiments, not all cache hints are implemented exactly to specification.


To confirm the behavior of caching hints available for DMA writes, 
we designed an AFU that writes a constant string to a configurable memory address using a configurable caching hint and bus.
We used the AFU to write a cache line and afterward timed a read access to the same cache line on the CPU.
These experiments confirm that nearly 100\% of the cache lines written to by the AFU are placed in the LLC, as access times 
stay below 100 CPU clock cycles while main memory accesses take 175 cycles on average. 
This behavior is independent of the caching hint, the bus, or the platform (PAC, integrated \textsf{Arria 10}).
The result is surprising as the caching hint meant to cache the data in the cache of the integrated \textsf{Arria 10} and 
the caching hint meant for writing directly to the main memory are either ignored by the Blue Region and the CPU or not implemented yet.
Intel later verified that the Blue Region in fact ignores all caching hints that apply to DMA writes. Instead, 
the CPU is configured to handle all DMA writes as if the \texttt{WrPush\_I} caching hint is set.
The observed LLC caching behavior is likely caused by Intel's Data Direct I/O (DDIO), which is enabled by default in recent Intel CPUs. DDIO is meant to give peripherals direct access to the LLC and thus causes the CPU to cache all memory lines written by the AFU.
DDIO restricts cache access to a subset of ways per cache set, which reduces the attack surface for Prime+Probe attacks. 
Nonetheless, attacks against other DDIO-enabled peripherals are possible~\cite{taram2019packetchasing,kurth2020netcat}.

\section{\attack\ Attack}\label{sec:Rowhammer}
\para{Contribution}
In this section, we present and evaluate a simple AFU for the \textsf{Arria 10 GX} FPGA that is capable of 
performing Rowhammer against its host CPU's DRAM as much as two times faster and four times more effectively than its host CPU can. 
In a Rowhammer, a significant factor in the speed and efficacy of an attack is the rate at which memory can be repeatedly accessed.
On many systems, the CPU is sufficiently fast to cause some bit flips, but 
the FPGA can repeatedly access its host machine's memory system substantially faster than the host machine's CPU can.
Both the CPU and FPGA share access to the same memory controller, 
but the CPU must flush the memory after each access to ensure 
that the next access reaches DRAM; memory reads from the FPGA do not affect the CPU cache system so no time 
is wasted flushing memory with the FPGA implementation.
We further measure the performance of CPU and FPGA Rowhammer implementations with caching both enabled and disabled, 
and find that disabling caching brings CPU Rowhammer speed near that of our FPGA Rowhammer implementation.
Crucially, the architectural difference also means that it is much more difficult for a program on the CPU 
to detect the presence of an FPGA Rowhammer than that of a CPU Rowhammer --- the FPGA's memory accesses leave far fewer traces on the CPU.

\subsection{\attack: Our FPGA Implementation of Rowhammer}
\attack\ supports configuration through the MMIO interface. When the \attack\ AFU is loaded,
the CPU first sets the target physical addresses that the AFU will repeatedly access. 
It is recommended to set both addresses for a double-sided attack, 
but if the second address is set to 0, \attack\ will perform a single-sided attack using just the first address. 
The CPU must also set the number of times to access the targeted addresses. 

When the configuration is set, the CPU signals the AFU to repeat memory accesses and issue them as fast as it can, 
alternating between addresses in a double-sided attack.
Note that unlike a software implementation of Rowhammer, the accessed addresses do not need 
to be flushed from cache --- DMA read requests from the FPGA do not cache the cache line in the CPU cache, 
though if the requested memory is in the last-level cache, the request will be served to the FPGA by the cache 
instead of by memory (see \autoref{sec:background_mem_cache_architecture} for more details on caching behavior).
In this attack, the attacker needs to ensure that the cache lines used for inducing bit flips 
are never accessed by the CPU during the attack.
The number of times to access the target addresses can be read again to get the number of remaining accesses; 
this is the simplest way to check in software whether or not the AFU has finished sending these accesses. 
When the last read request has been sent by the AFU, 
the total amount of time taken to send all of the requests is recorded.\footnote{
The time to send all the requests is not precisely the time to complete all the requests, 
but it is very close for sufficiently high numbers of requests. The FPGA has a transaction buffer that holds up to 64 transactions 
after they have been sent by the AFU. The buffer does take some time to clear, but the additional time is negligible for our performance measurements of millions of requests.}

\subsection{\attack{} on the FPGA PAC vs. CPU Rowhammer}
\begin{figure}[tb]\centering
 {\includegraphics[width=\linewidth]{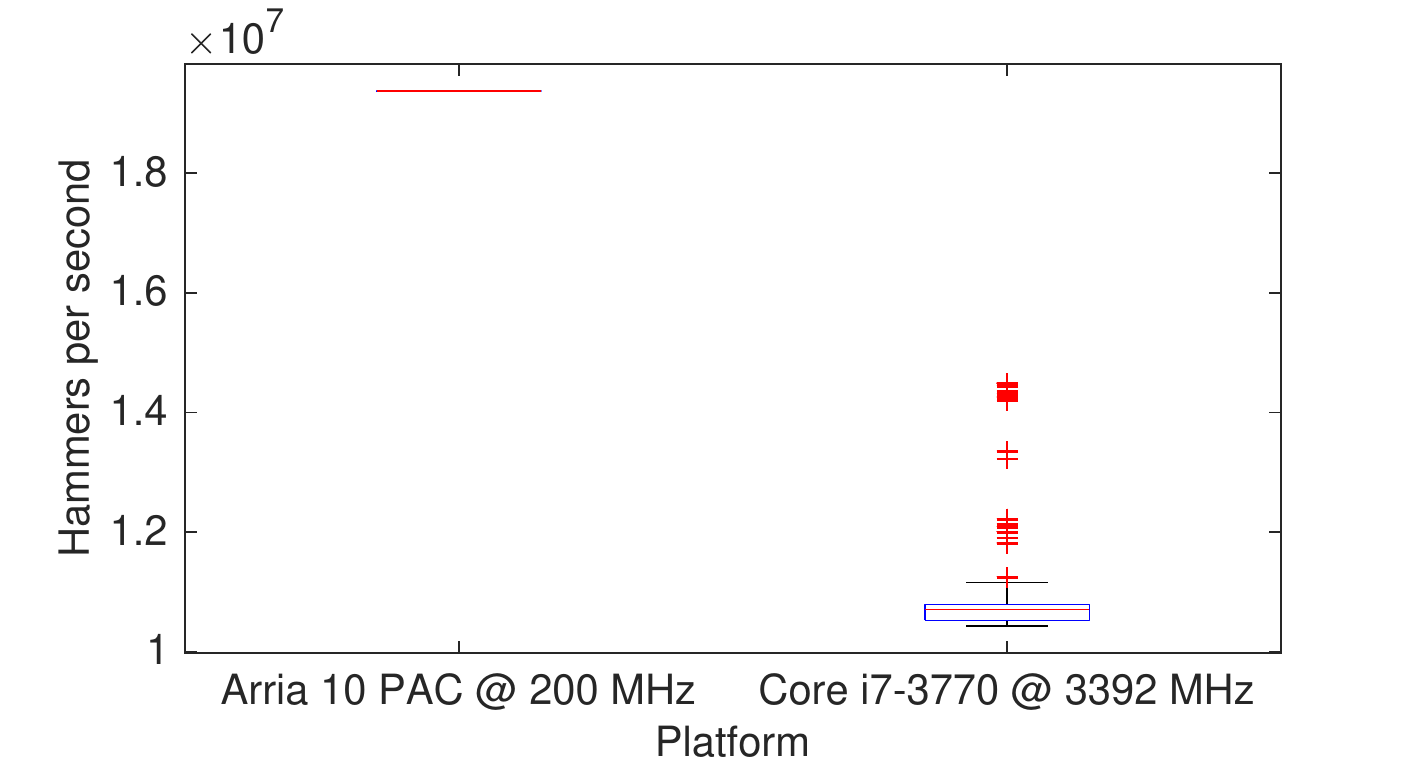}}
 {\caption{Distributions of hammering rates (memory requests per second) on FPGA PAC and i7-3770.}
 \label{fig:pac_hammer_rate}}
\end{figure}

\autoref{fig:pac_hammer_rate} shows a box plot of the 0th, 25th, 50th, 75th, and 100th percentile of 
measured ``hammering rates'' on the \textsf{Arria 10} FPGA PAC and its host i7-3770 CPU. 
Each measurement in these distributions is the average hammering rate over a run of 2 billion memory requests. 
Our \attack{} implementation is substantially faster than the standard CPU Rowhammer, 
and its speed is far more consistent than the CPU's. 
The FPGA can manage an average throughput of one memory request, or ``hammer,'' every ten \unit[200]{MHz} 
FPGA clock cycles (finishing 2 billion hammers in an average of 103.25 seconds); 
the CPU averages one hammer every 311 \unit[3.4]{GHz} CPU clock cycles 
(finishing 2 billion hammers in an average of 183.41 seconds). 
Here we can see that even if the FPGA were clocked higher, it would still spend most of its 
time waiting for entries in the PCIe transaction buffer in the non-reconfigurable region to become available. 

\begin{figure}[tb]\centering
 {\includegraphics[width=\linewidth]{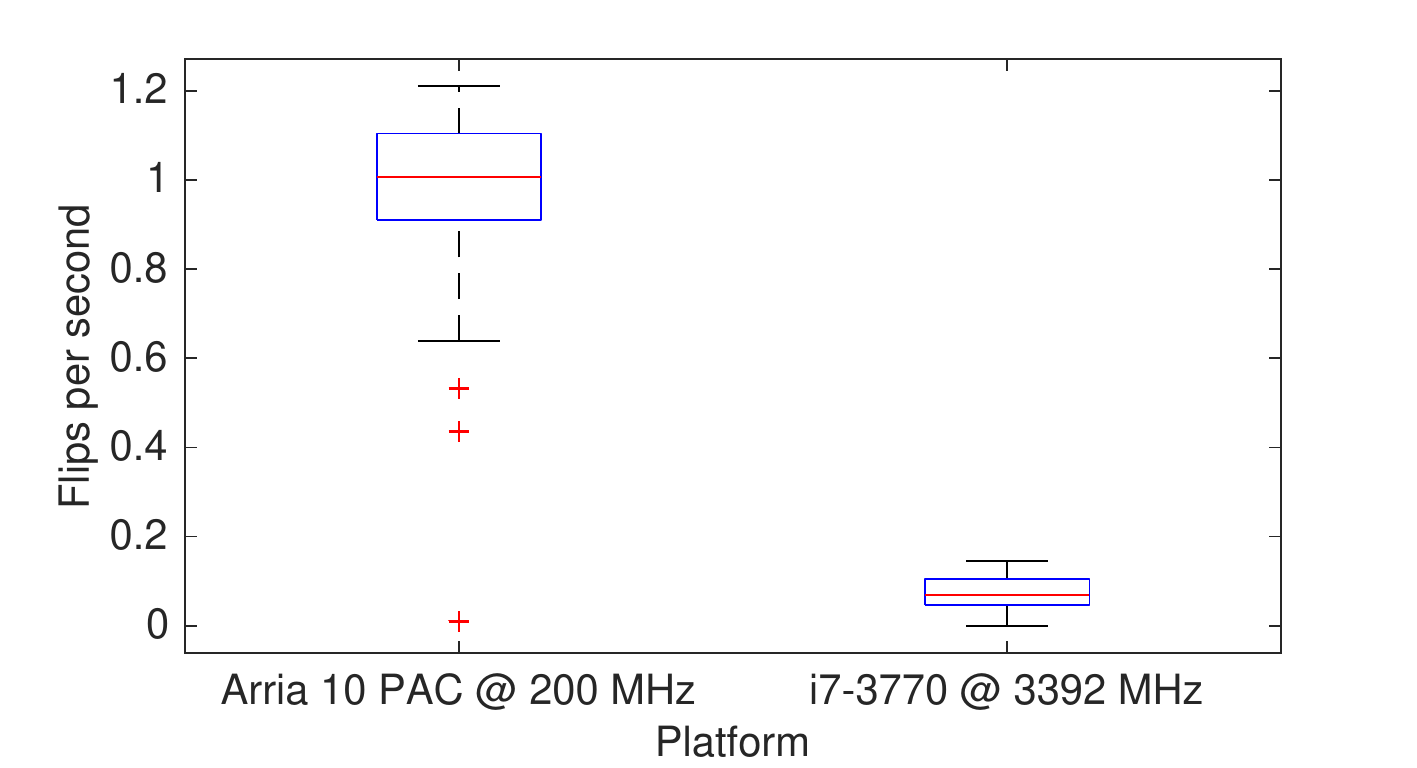}}
 {\caption{Distributions of flip rates on FPGA PAC and i7-3770.}
 \label{fig:pac_flip_rate}}
\end{figure}

\autoref{fig:pac_flip_rate} shows measured bit flip rates in the victim row for the same experiment.
Runs where zero flips occurred during hardware or software hammering were excluded from the flip rate distributions, 
as they are assumed to correspond with sets of rows that are in the same logical bank, but not directly adjacent to each other. 
The increased hammering speed of \attack\ produces a more than proportional increase in flip rate, 
which is unsurprising due to the highly nature of Rowhammer.
As the Rowhammer is underway, electrical charge is drained from capacitors in the victim row.
However, the memory controller also periodically refreshes the charge in the capacitors.
When there are more memory accesses to adjacent rows within each refresh window, 
it is more likely that a bit flip occurs before the next refresh.
This is why the FPGA's increased memory throughput is more effective for conducting 
Rowhammer against the same DRAM chip.
\begin{figure}[tb]
 \includegraphics[width=\linewidth]{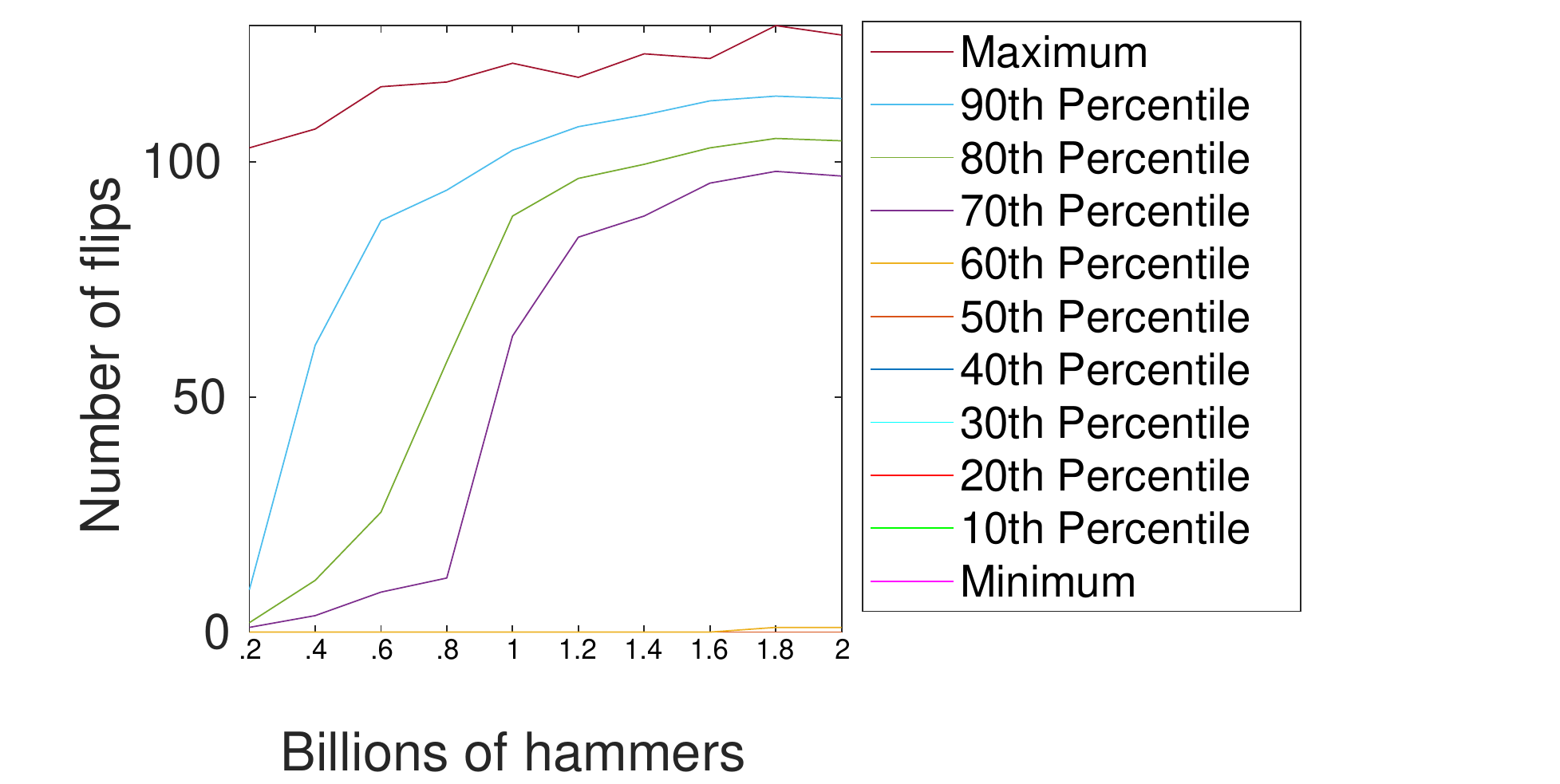}
  \caption{Distributions of total flips after 200 million to 2 billion hammers on PAC.}
 \label{fig:quant_flips_hw}
\end{figure}

\begin{figure}[tbp]\centering
 {\includegraphics[width=\linewidth]{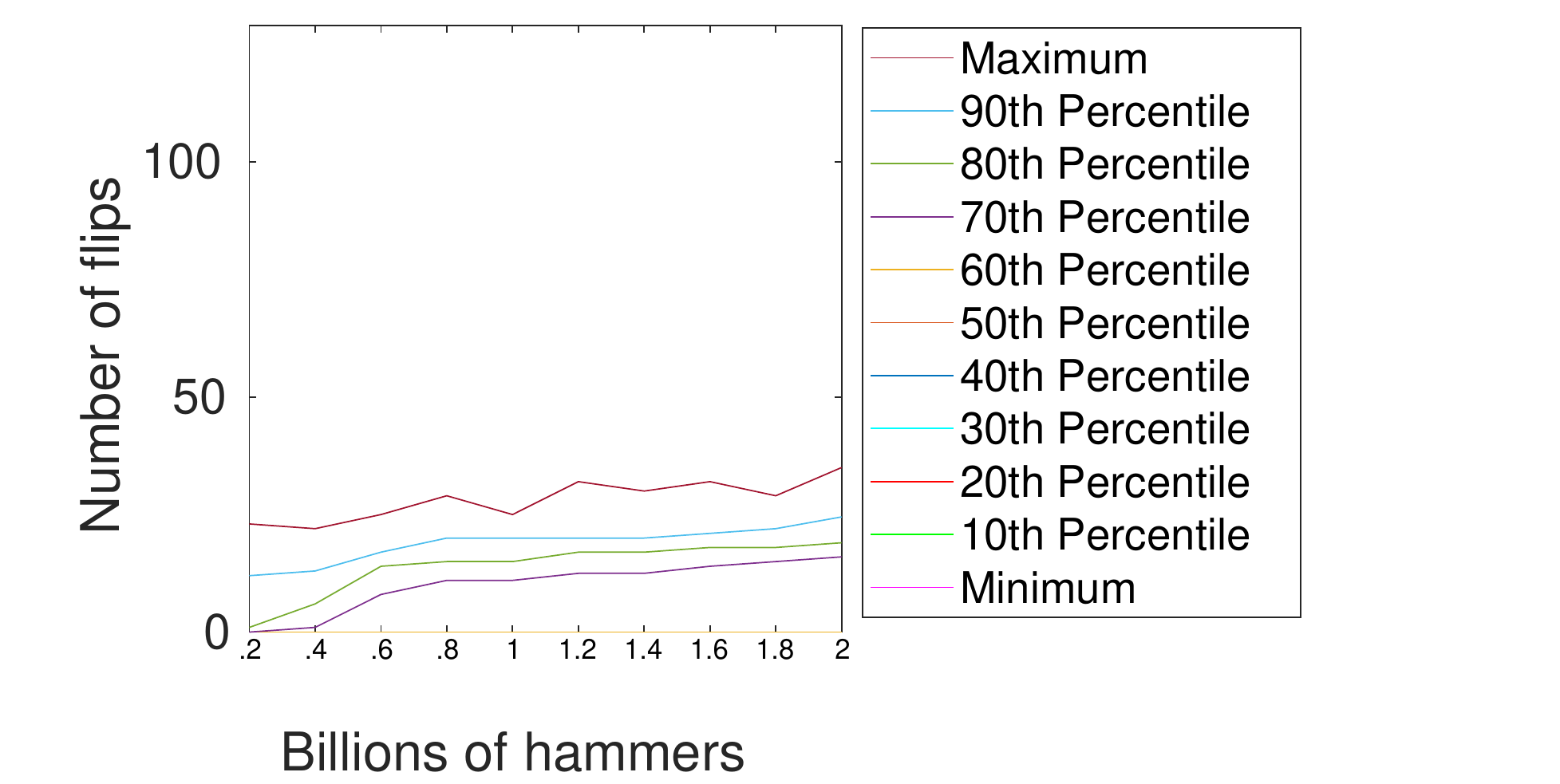}}
 {\caption{Distributions of total flips after 200 million to 2 billion hammers on i7-3770.}
 \label{fig:quant_flips_sw}}
\end{figure}

Another way to look at hammering performance is by counting the total number of flips produced by a given number of hammers. 
\autoref{fig:quant_flips_hw} and \autoref{fig:quant_flips_sw} show minimum, maximum, and every 10th percentile of the number 
of flips produced by the AFU and CPU respectively for a range of total number of hammers from 200 million to 2 billion.
These graphs demonstrate how much more effectively the FPGA PAC can generate bit flips in the DRAM even after the same number of memory accesses. 
For hammering attempts that resulted in a non-zero number of bit flips, 
the AFU exhibits a wide distribution of flip count in the range of 200 million to 800 million hammers which then 
rapidly narrows in the range of 800 million to 1.2 billion and finally levels out by 1.8 billion hammers. 
This set of distributions seems to indicate that ``flippable'' rows will ultimately reach about 80-120 total flips after enough hammering, 
but it can take anywhere from 200 million hammers (about 10 seconds) to 2 billion hammers (about 100 seconds) to reach that limit.

\begin{figure}[tb]
 \centering
 \includegraphics[width=1\linewidth]{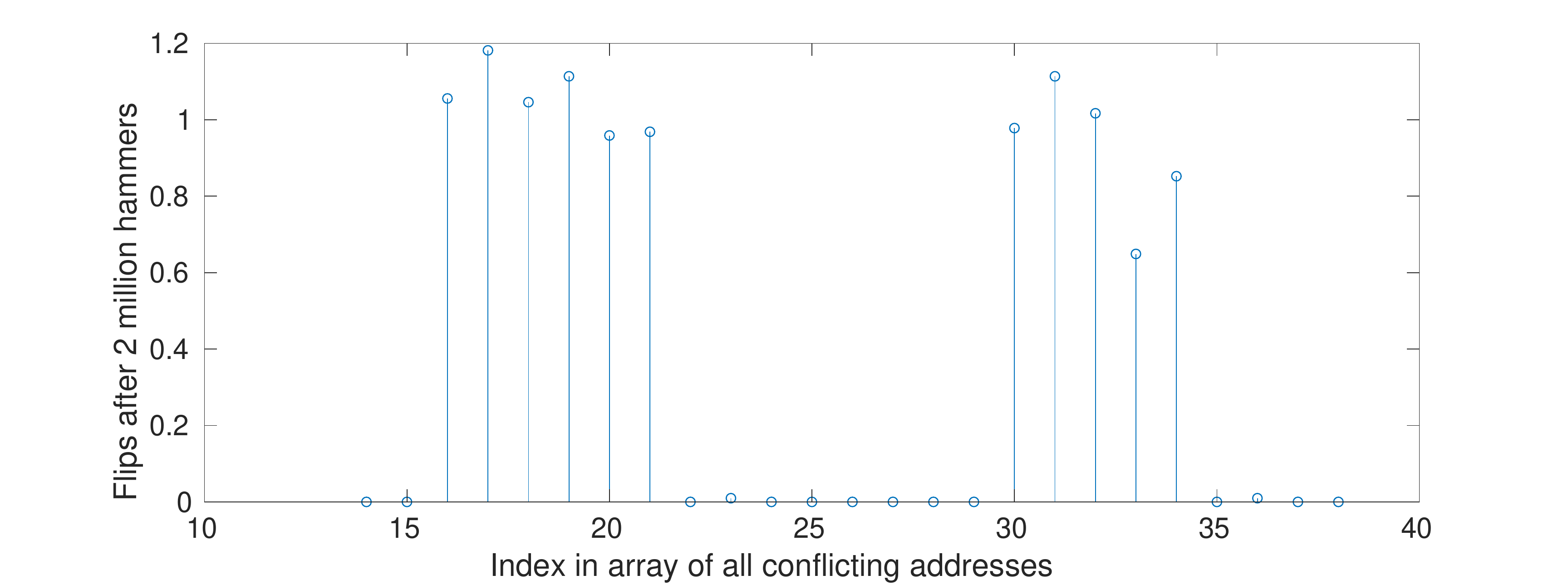} 
 \caption{Time series plotting number of flips on a row-by-row basis, showing an example of the consistent placement of small-valued outliers (samples 23 and 36 on this graph) relative to their much larger neighbors. These rows only ever incur a few flips, compared to most ``flippy'' rows which incur dozens of flips, and always are located two rows away from a block of rows that flip much more.}
  \label{fig:outlier_impulses}
\end{figure}

There are also a few rows that only incur a few flips. 
These samples appear in a consistent pattern demonstrated in \autoref{fig:outlier_impulses}, 
which plots a portion of the data used to create \autoref{fig:quant_flips_hw} in detail.\zane{this figure is a good candidate for the appendix if we go down that road}
Each impulse in this plot represents the number of flips after a single run of 2 billion hammers on a particular target row. 
In \autoref{fig:outlier_impulses}, at indices 23 and 36, two of these outliers are visible, each appearing two indices after several 
samples in the standard 80-120 flip range. These outliers could indicate rows that are affected vary slightly by hammering 
on rows that are nearby but not adjacent.

 \subsection{\attack\ on the \textsf{Integrated Arria 10} vs. CPU Rowhammer}
The \attack{} AFU we designed for the integrated platform is the same as the AFU for the PAC, 
except that the integrated platform has access to more physical channels for the memory reads. 
The PAC only has a single PCIe channel; the integrated platform has one UPI channel and two PCIe channels, 
as well as an ``automatic'' setting which lets the interface manager select a physical channel automatically. 
Therefore we present the hammering rates on this platform with two different settings --- alternating PCIe lanes on each access and 
using the automatic setting.

\begin{figure}[tbp]\centering
  {\includegraphics[width=\linewidth]{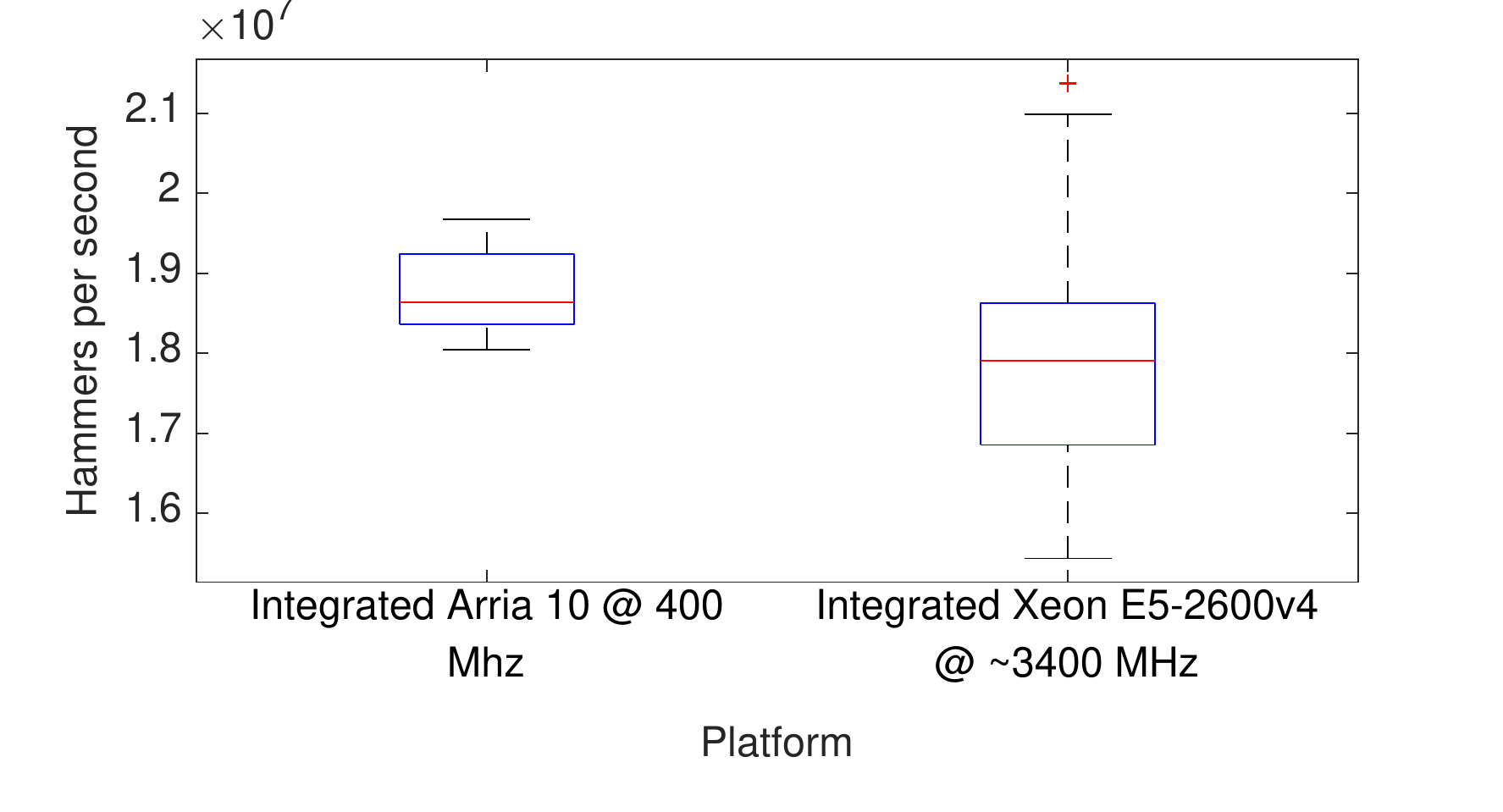}}
  {\caption{Distributions of hammering rates on integrated \textsf{Arria 10} and Xeon E5-2600 v4.}
  \label{fig:int_hammer_rate}}
\end{figure}

However, this platform is only available to us on Intel's servers, so we have only been able to test on one DRAM setup 
and have been unable to get bit flips on this DRAM.\footnote{
There are several reasons why this could be the case. 
Some DRAM is simply more resistant to Rowhammer by its physical nature. 
DDR4 memory, which can be found in this system, sometimes has hardware features to block Rowhammer style attacks~\cite{jedec};
though some methods have been developed to circumvent these protections~\cite{gruss2018another}, 
these methods ultimately still rely on the ability of the attacker to repeatedly access the DRAM very quickly, 
so we consider those methods outside of the scope of this research, which is focused on the relative ability of the FPGA platforms 
to quickly access DRAM.}
The integrated \textsf{Arria 10} shares the package with a modified Xeon v4-style CPU.
The available servers are equipped with an X99 series motherboard with \unit[64]{GB} of DDR4 memory. 
\autoref{fig:int_hammer_rate} shows distributions of measured hammering rates on the integrated \textsf{Arria 10} platform. 
Compared to the \textsf{Arria 10} PAC, the integrated \textsf{Arria 10}'s hammering rate is more varied, but with a similar mean rate. 

\begin{figure}[tbp]\centering
  {\includegraphics[width=\linewidth]{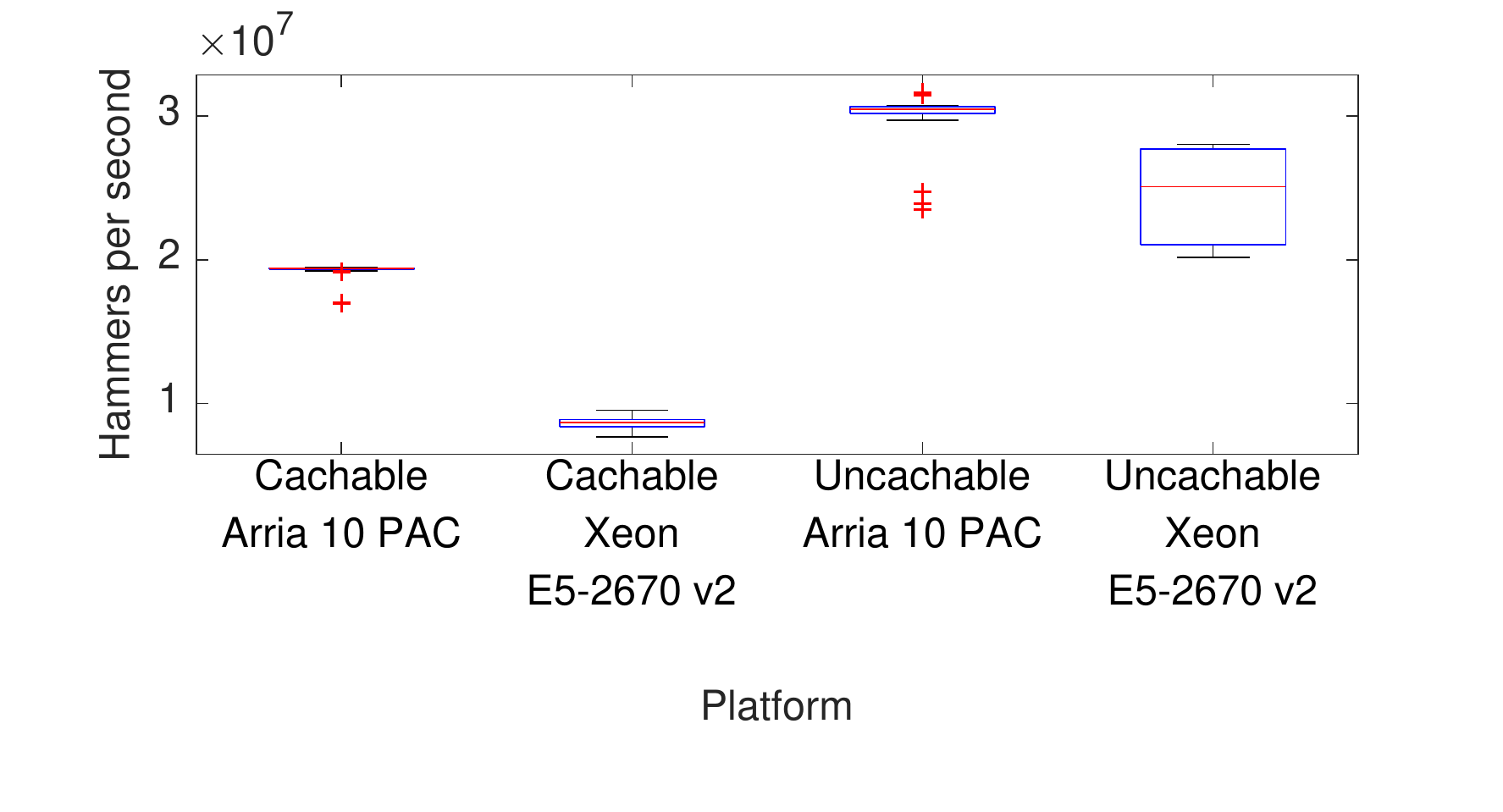}}
  {\caption{Distributions of hammering rates with cachable and uncachable memory.}
  \label{fig:cache_hammer_rate}}
\end{figure}

\subsection{The Effect of Caching on Rowhammer Performance} 
We hypothesized that a primary reason for the difference in Rowhammer performance between \attack{} 
on the FPGAs and a typical Rowhammer implementation on the CPUs is that when one of the FPGAs reads 
a line of memory from DRAM, it is not cached, so the next read will miss the cache and be directed to the DRAM as well. 
On the other hand, when the CPUs access a line of memory, it is cached, and the memory line must be flushed 
from cache before the next read is issued, or the next read will hit the cache instead of DRAM, 
and the physical row in the DRAM will not be ``hammered.'' 

To evaluate our hypothesis that caching is an important factor in the performance disparity we observed between FPGA- and CPU-based Rowhammer, 
we used the PTEditor~\cite{schwarz2019pteditor} kernel module to set allocated pages as uncachable before testing hammering performance.
We edited the setup of the Rowhammer performance tests to allocate many \unit[4]{kB} pages and set all of those as uncachable 
instead of one \unit[2]{MB} huge page, as the kernel module we used to set the pages as uncachable was not correctly 
configuring the huge pages as uncachable.
However, it is still easy to find a large continuous range of physical addresses --- when these pages are 
allocated by \textsf{OPAE}, the physical address is directly available to the software. 
So the software simply allocates thousands of \unit[4]{kB} pages, sorts them, and then finds the biggest continuous 
range within them and attempts to find colliding row addresses within that range. 
The \attack{} AFU required no modifications from the initial performance tests; the assembly code to hammer from the CPU 
was edited to not flush the memory after reading it, since the memory will not be cached in the first place. 

We performed this experiment by placing the FPGA PAC on a Dell Poweredge R720 system with a Xeon E5-2670 v2 CPU fixed to a clock speed of 
\unit[2500]{MHz} and two \unit[4]{GB} DIMMs of DDR3 DRAM clocked at \unit[1600]{MHz}. 
\autoref{fig:cache_hammer_rate} shows the performance of the FPGA PAC and this system's CPU with caching enabled and disabled.
Disabling caching produces a significant speedup in hammering for both the PAC and the CPU, 
but especially for the CPU, which saw a 188\% performance increase. 
With caching enabled, the median hammering rate of the PAC was more than twice that of the CPU, 
but with caching disabled, the median hammering rate of the PAC was only 22\% faster than that of the CPU. 
Of course, memory accesses on modern systems are extremely complex (even with caching disabled), 
so there are likely some factors affecting the changes in hammering rate that we cannot describe, 
but our experimental evidence supports our hypothesis that time spent flushing the cache is a major factor 
slowing down CPU Rowhammer implementations compared to FPGA implementations.

\section{Fault Attack on RSA using \attack}\label{sec:rsa_fault_Rowhammer}
Rowhammer has been used for fault injections on cryptographic schemes~\cite{bhattacharya2016curious,bhattacharya2018advanced} 
or for privilege escalation~\cite{gruss2018another,veen2016drammer,seaborn2015exploiting}. 
Using \attack, we demonstrate a practical fault injection attack from the \textsf{Arria 10} FPGA
to the \textsf{WolfSSL} RSA implementation running on its host CPU. 
In the RSA fault injection attack proposed by Boneh et al.~\cite{boneh1997importance}, 
an intermediate value in the Chinese remainder theorem modular exponentiation algorithm is faulted, 
causing an invalid signature to be produced. 
Similarly, we attack the \textsf{WolfSSL} RSA implementation using \attack\ from the FPGA PAC and Rowhammer from the host CPU,
and compare the efficiency of the two attacks. 
The increased hammering speed and flip rate of the \textsf{Arria 10} FPGA 
makes the attack more practical in the time frame of about 9 RSA signatures.

\begin{figure*}[tb]
 \centering
 \includegraphics[width=\linewidth]{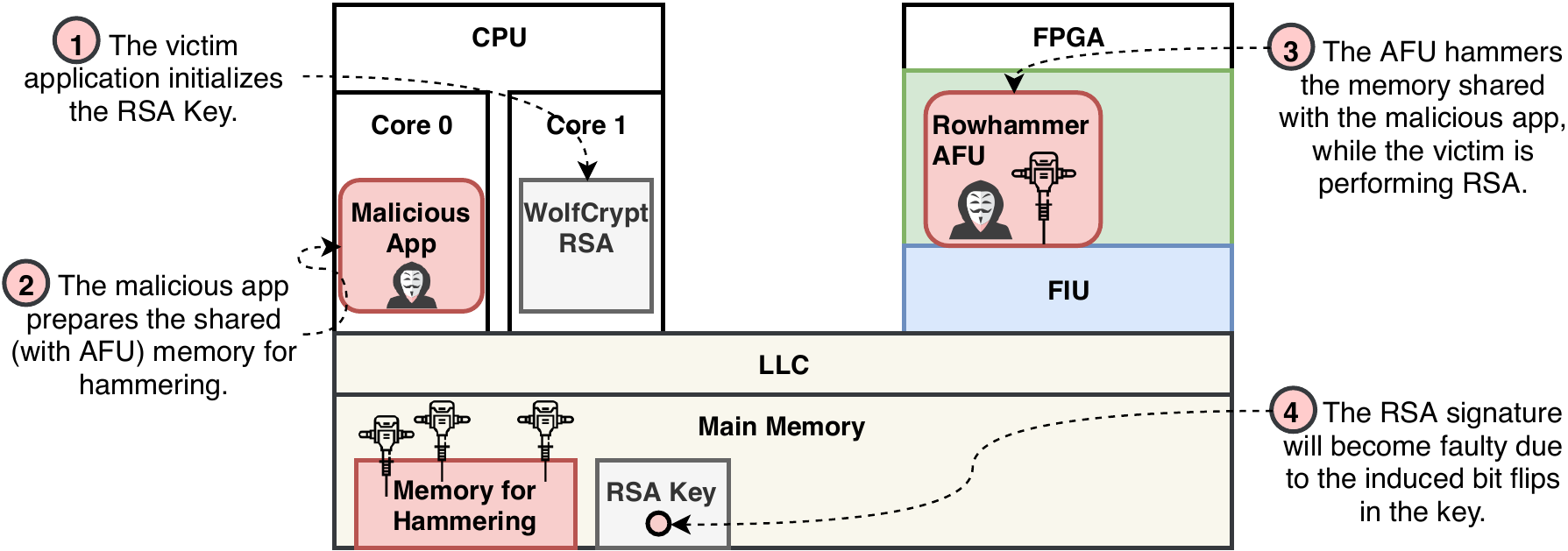}
 \caption{\textsf{WolfSSL} RSA Fault Injection Attack.}
 \label{fig:fault_injection}
\end{figure*}

\autoref{fig:fault_injection} shows the high-level overview of our attack: 
the \textsf{WolfSSL} RSA application runs on one core, while a malicious application runs adjacent to it, 
assisting the \attack\ AFU on the FPGA in setting up the attack. 
\attack\ causes a hardware fault in the main memory, and 
when the \textsf{WolfSSL} application reads the faulty memory, 
it produces a faulty signature and leaks the private factors used in the RSA scheme.

\subsection{RSA Fault Injection Attacks}\label{sec:rsa_algs}
We implement a fault injection attack against the Chinese remainder theorem 
implementation of the RSA algorithm, commonly known as the Bellcore attack~\cite{boneh1997importance}. 
\autoref{alg:rsa-crt} shows the Chinese remainder theorem (CRT) RSA signing scheme 
where the signature $S$ is computed by raising a message $m$ to the private exponent 
$d$th power, modulo $N$. $d_p$ and $d_q$, are precomputed as $d \bmod p-1$ and $d \bmod q-1$, 
where $p$ and $q$ are the prime factors of $N$~\cite{aumuller2002fault}.
When one of the intermediates $S_q$ or $S_p$ is computed incorrectly, an interesting case arises. 
Consider the difference between a correctly computed signature $S$ of a message $m$ and an 
incorrectly computed signature $S'$ of the same message, computed with an invalid intermediate $S'_p$. 
The difference $S-S'$ leaves a factor of $q$ times the difference $S_p-S'_p$, so the GCD of $S-S'$ and $N$ is 
the other factor $p$~\cite{aumuller2002fault}. 
This reduces the problem of factoring $N$ to a simple subtraction and GCD operation, 
so the private factors $(p,q)$ are revealed if the attacker has just one valid signature and one faulty signature, each signed on the same message $m$. 
These factors can also be recovered with just one faulty signature 
if the message $m$ and public key $e$ are known; it is also equal to the GCD of $S'^e-m$ and $N$.

\para{Fault Injection Attack with RSA Base Blinding}
A common modification to any RSA scheme is the addition of base blinding, 
effective against simple and differential power analysis side-channel attacks, 
but vulnerable to a correlational power analysis attack demonstrated by~\cite{witteman2011defeating}. 
Base blinding is used by default in our target \textsf{WolfSSL} RSA-CRT signature scheme.
In this blinding process, the message $m$ is blinded by a randomly generated number $r$ by computing $m_b=m\cdot r^e \bmod n$. 
The resulting signature $S_b=(m \cdot r^e)^d \bmod n=m^d \cdot r \bmod n$ must then be multiplied by the inverse of the random number $r$ 
to generate a valid signature $S=S_b \cdot r^{-1} \bmod n$. 



This blinding scheme does not prevent against the Bellcore fault injection attack. 
Consider a valid signature blinded with random factor $r_1$ and an invalid signature blinded with $r_2$. 
When the faulty signature is subtracted by the valid signature, 
the valid and blinded intermediates $S_{pb}$ are each unblinded and cancel as before, 
as shown in \autoref{eq:blind_fault_1}.
\begin{equation}
  \resizebox{\linewidth}{!}{$
\begin{split}
\label{eq:blind_fault_1}
 S-S' = &\left[S_{qb} + q\cdot \left((S_{pb}-S_{qb})\cdot q^{-1}\bmod p\right)\right]\cdot r_1^{-1} \bmod N \\
 &- \left[S_{qb} + q\cdot \left((S'_{pb}-S_{qb})\cdot q^{-1}\bmod p\right)\right]\cdot r_2^{-1} \bmod N\\
 = &\ q\cdot [\left(S_{pb}\cdot q^{-1}\bmod p\right)\cdot r_1^{-1} - \left(S'_{pb}\cdot q^{-1}\bmod p\right)\cdot r_2^{-1}] \bmod N
\end{split}
$}
\end{equation}
Ultimately, there is still a factor of $q$ in the the difference $S-S'$ which can be extracted with a GCD as before.

\subsection{Our Attack}
\para{Approach and Justification} 
We developed a simplified attack model to test the effectiveness of the \textsf{Arria 10} Rowhammer in a fault injection scenario.
Our model simplifies the setup of the attack so that we can efficiently measure the performance of both CPU Rowhammer and \attack{}.
We sign the same message with the same key repeatedly while the Rowhammer exploit runs, 
and count the number of correct signatures until a faulty signature is generated, which is used to leak the private RSA key.

\para{Attack Setup}
In summary, our simplified attack model works as follows: 
The attacker first allocates a large block of memory and checks it for conflicting row addresses.
It then quickly tests which of those rows can be faulted with hammering using \attack.
A list of rows that incur flips is saved so that it can be iterated over.
The program then begins the ``attack,'' iterating through each row that incurred flips during the test, 
and through the sixty-four 1024-bit offsets that make up the row.
During the attack, the \attack\ AFU is instructed to repeatedly access the rows adjacent to the target row.
Meanwhile, in the ``victim'' program, the targeted data (the precomputed intermediate value $d \bmod q-1$) 
is copied to the target address, which is computed as an offset of the targeted row.
The victim then enters a loop where it reads back the data from the target row and uses it as part of an RSA key 
to create a signature from a sample message.
Additionally, the ``attacker'' opens a new thread on the CPU which repeatedly flushes the target row on a given interval.
It is necessary for the attacker to flush the target row because the victim is repeatedly reading the targeted data and 
placing it in cache, but the fault will only ever occur in main memory. 
For the victim program to read the faulty data from DRAM, there cannot be an unaffected copy of the same data in cache 
or the CPU will simply read that copy.
As we show below, the performance of the attack depends significantly on the time interval between flushes. 

One of the typical complications of a Rowhammer fault injection attack is ensuring that the victim's data 
is located in a row that can be hammered.
In our simplified model, we choose the location of the victim data manually within a row that 
we have already determined to be one that incurs flips under a Rowhammer attack so that we may easily test the 
effectiveness of the attack at various rows and various offsets within the rows.
In a real attack, the location of the victim program's memory can be controlled by the attacker with a technique 
known as page spraying~\cite{seaborn2015exploiting}, which is simply allocating a large number of pages and then 
deallocating a select few, filling the memory in an attempt to cause the victim program to allocate the right pages.
Improvements in this process can be made; for example,~\cite{bhattacharya2016curious} demonstrated how cache 
attacks can be used to gather information about the physical addresses of data being used by the victim process. 

The other simplification in our model is that we force the CPU to read from DRAM using 
the \texttt{clflush} instruction to flush the targeted memory from cache.
In an end-to-end attack, the attacker would use an eviction set to evict the targeted memory 
since it is not directly accessible in the attack process's address space.
However, the effect is ultimately the same --- the targeted data is forcibly removed from the cache by the attacker.

\subsection{Performance of the Attack}
In this section, we show that our \attack\ implementation with optimal settings can cause a faulty signature 
an average of 17\% faster than a typical CPU-based, software-driven Rowhammer implementation with optimal settings. 
In some scenarios, the performance is as much as 4.8 times that of the software implementation.
However, under some conditions, the software implementation can be more likely to cause a fault over a longer period of time.
Our results indicate that increasing the DRAM row refresh rate provides significant but not complete 
defense against both implementations.

The performance of this fault injection attack is highly dependent on the time interval between evictions, 
and as such we present all of our results in this section as functions of the eviction interval.
Each eviction triggers a subsequent reload from memory when the key is read for the next signature, 
which refreshes the capacitors in the DRAM.
Whenever DRAM capacitors are refreshed, any accumulated voltage error in each capacitor 
(due to Rowhammer or any other physical effect) is either solidified as a new faulty bit value or reset to a safe and correct value.
Too short of an interval between evictions will cause the DRAM capacitors to be refreshed too quickly to be flipped with a high probability. 
On the other hand, however, longer intervals can mean the attack is waiting to evict the memory
 for a longer time while a bit flip has already occurred.
It is crucial to note, also, that DRAM capacitors are automatically refreshed by the memory controller on a 64 ms interval\footnote{
  More specifically, DDR3 and DDR4 specifications indicate \unit[64]{ms} as the maximum allowable time between DRAM row refreshes.
}
~\cite{gruss2016Rowhammer}.
On some systems, this interval is configurable: faster refresh rates reduce the rate of memory errors, 
including those induced by Rowhammer, but they can impede maximum performance because the memory spends 
more time doing maintenance refreshes rather than serving read and write request.
For more discussion on modifying row refresh rates as a defense against Rowhammer, see \autoref{sec:countermeasures}.

\begin{figure}[tb]
 \centering
 \includegraphics[width=1\linewidth]{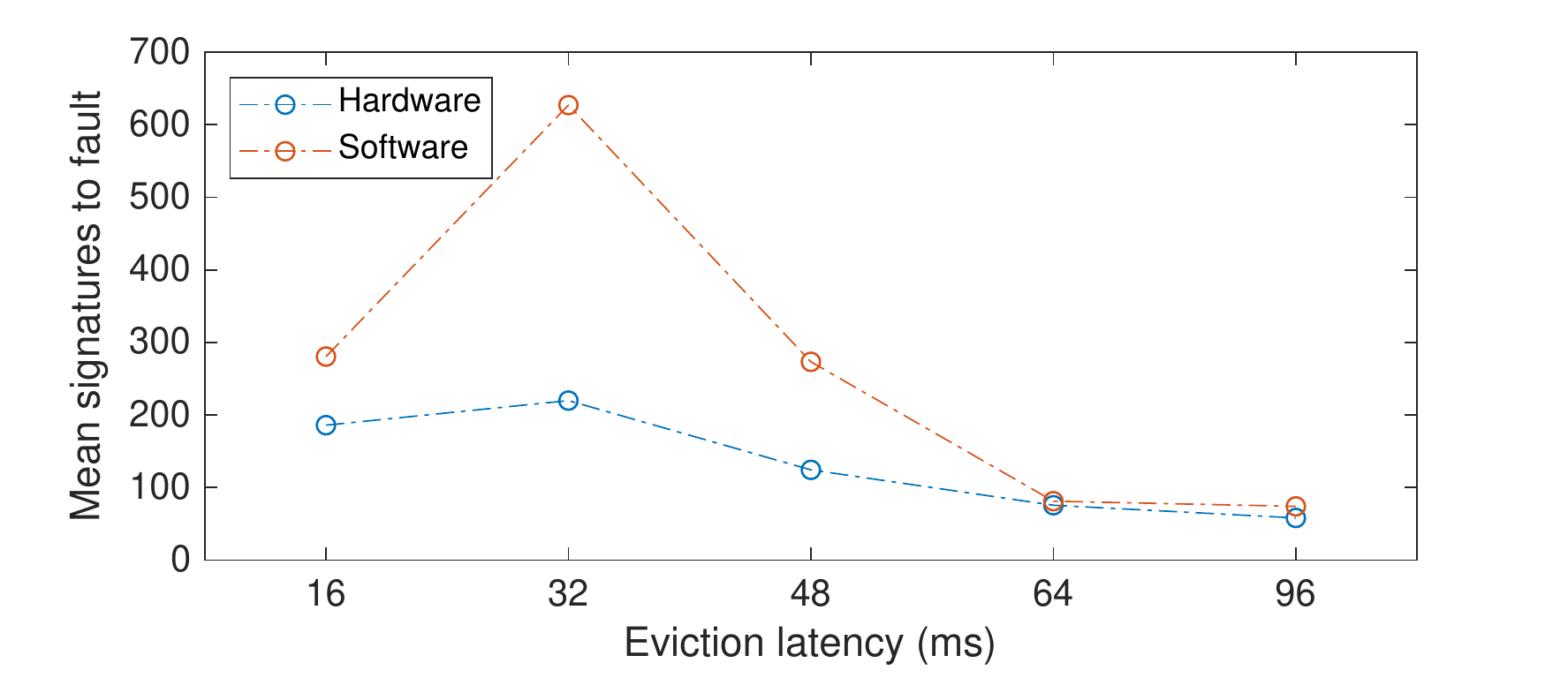}
 \caption{Mean number of signatures to fault at various eviction intervals.}
 \label{fig:attack_performance}
\end{figure}

In \autoref{tab:performance} we present two metrics with which we compare \attack{} and a standard CPU Rowhammer implementation.
This table shows the mean number of signatures until a faulty signature is produced and the ultimate probability 
of success of an attack within 1000 signatures against a random key in a randomly selected chunk of memory 
within a row known to be vulnerable to Rowhammer.
With an eviction interval of \unit[96]{ms}, the \attack{} attack achieves the lowest average number of 
signatures before a fault, at only 58, 25\% faster than the best performance of the CPU Rowhammer.
The CPU attack is impeded significantly by shorter eviction latency, while the \attack{} implementation is not, 
indicating that on systems where the DRAM row refresh rate has been increased to protect against memory faults and Rowhammers, 
\attack{} likely offers substantially improved attack performance.
\autoref{fig:attack_performance} highlights the mean number of signatures until a faulty signature for the \unit[16]{ms} to \unit[96]{ms} range of eviction latency.

\begin{table*}[tb]
  \centering
  \caption{Performance of our \attack{} exploit compared to a standard software CPU Rowhammer with various eviction intervals. \attack{} is able to acheive better performance in many cases because it bypasses caching architecture, sending more memory requests during the eviction interval and causing bit flips at a higher rate.}
  \label{tab:performance}
  \resizebox{.8\hsize}{!}{
  \begin{tabular}{r|c c c c c c c}
    \toprule
     Eviction & \multicolumn{3}{c}{Mean signatures to fault}&\ & \multicolumn{3}{c}{Successful fault rate}\\
    \cline{2-4} \cline{6-8}
    Interval & CPU & \attack\ & \% Inc. Speed &\ & CPU & \attack\ & \% Inc. Rate \\
    \midrule
    \rowcolor{gray!25}
    16  & 280 & 186 &  51\% & & 0.4\%  &0.2\% & -46\% \\
    32  & 627 & 219 & 185\% & & 0.2\%  &0.8\% & 264\% \\
    \rowcolor{gray!25}
    48  & 273 & 124 & 120\% & &  14\%  & 19\% &  39\% \\
    64  &  81 &  76 &   7\% & &  17\%  & 26\% &  56\% \\
    \rowcolor{gray!25}
    96  &  74 &  58 &  27\% & &  46\%  & 49\% &   8\% \\
    128 &  73 &  70 &   4\% & &  52\%  & 50\% &-1.2\% \\
    \rowcolor{gray!25}
    256 & 106 & 115 &  -7\% & &  57\%  & 55\% &  -3\% \\
    {\bfseries Best performance}
        &  {\bfseries 73} &  {\bfseries 58} &  {\bfseries 25\%} & &  {\bfseries 57\%}  & {\bfseries 55\%} &{\bfseries -3\%}\\
    \bottomrule
  \end{tabular}    
  }
\end{table*}

\vspace{-10pt}
\section{Cache Attacks on Intel FPGA-CPU Platforms}
\label{sec:side_channel}
In \autoref{sec:caching_hints}, we reverse-engineered the behavior of the memory subsystem 
on current \textsf{Arria 10} based FPGA-CPU platforms. 
In this section, we systematically analyze cache attacks exploitable by an AFU- or CPU-based attacker attacking the CPU or FPGA, respectively, and demonstrate a cache covert channel from FPGA to CPU.
At last, we discuss the viability of intra-FPGA cache attacks. \autoref{tab:cacheattackfindings} summarizes our findings.

\begin{table}[b]
	\centering
  \caption{Summary of our	cache attacks analysis: \textsf{OPAE} accelerates eviction set construction by making huge pages and physical addresses available to userspace.}
  \label{tab:cacheattackfindings}
  \resizebox{\linewidth}{!}{
	\begin{tabular}{cccc}
		\toprule
		\small{Attacker}            & \small{Target}     & \small{Channel} & \small{Attack} \\
		\midrule
		{FPGA PAC AFU}        & {CPU LLC}    & {PCIe  }  & {E+T, E+R, P+P}\\
		{Integrated FPGA AFU} & {CPU LLC}    & {UPI   }  & {E+T, E+R, P+P} \\
		{Integrated FPGA AFU} & {CPU LLC}    & {PCIe  }  & {E+T, E+R, P+P} \\
		{CPU}                 & {FPGA Cache} & {UPI   }  & {F+R, F+F} \\
		{Integrated FPGA AFU} & {FPGA Cache} & {CCI-P}   & {E+T, E+R, P+P} \\
		\bottomrule
  \end{tabular}
  }
\end{table}

To measure memory access latency on the FPGA, we designed a timer clocked at \unit[200]{MHz}/\unit[400]{MHz}.
The advantage of this hardware timer is that it runs uninterruptible in parallel to all other CPU or FPGA operations. 
Therefore, the timer precisely counts FPGA clock cycles, while timers on the CPU, such as \texttt{rdtsc}, 
may yield noisier measurements due to interruptions by the OS and the CPU's out-of-order pipeline. 

\subsection{Cache Attacks from FPGA PAC to CPU}
\begin{figure}[tb]
	\centering
			\includegraphics[width=\linewidth]{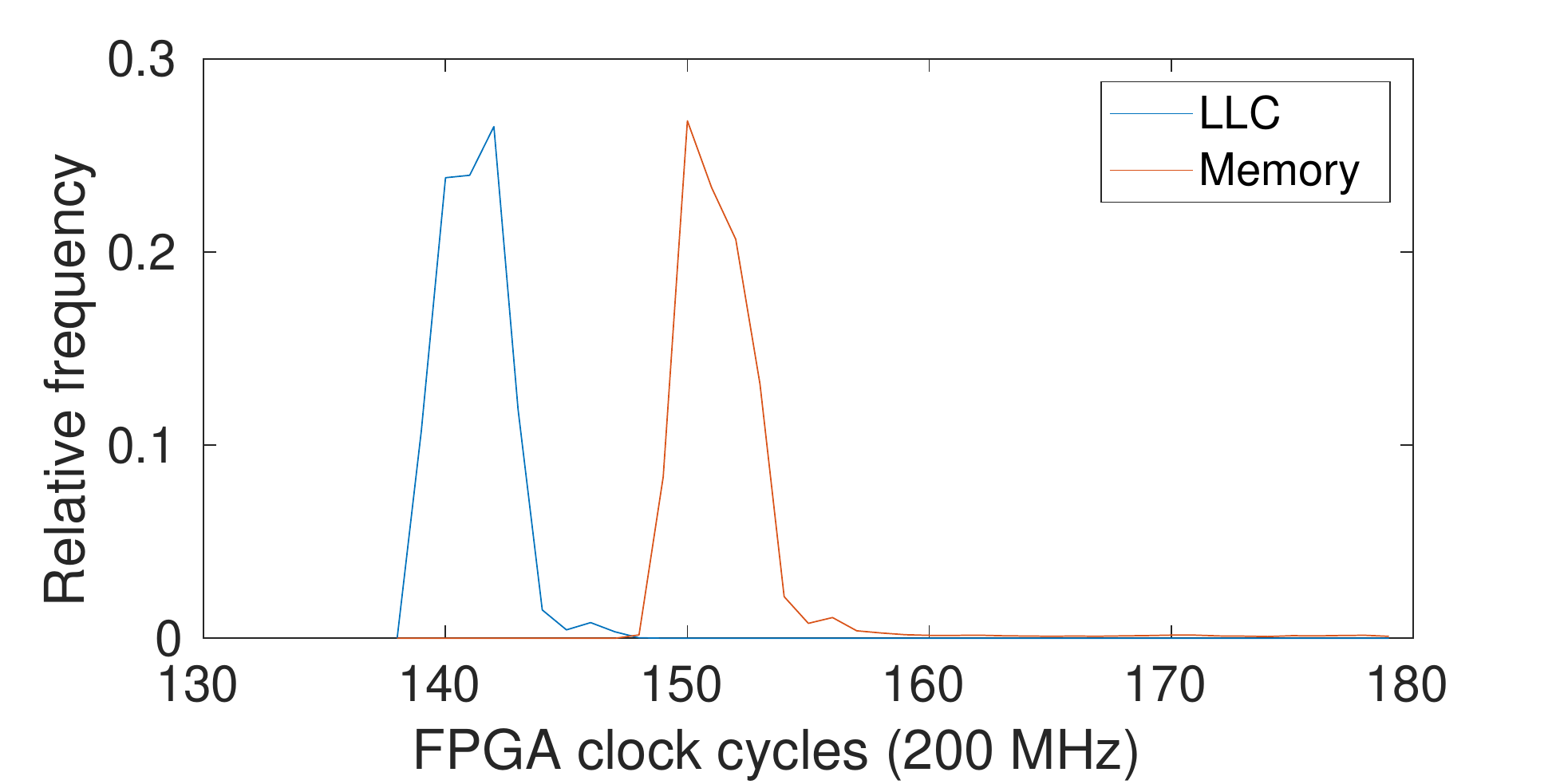}
		}{
			\caption{Latency for PCIe read requests on an FPGA PAC served by the CPU's LLC or main memory.}
			\label{fig:pac_llc_hit_miss}
    \end{figure}

The Intel PAC has access to one PCIe lane that connects it to the main memory of the system through the CPU's LLC. 
The \textsf{CCI-P} documentation~\cite{intel2018ccip} mentions a timing difference for memory requests served by 
the CPU's LLC and those served by the main memory. 
Using our timer we verified the suggested differences as shown in \autoref{fig:pac_llc_hit_miss}. 
Accesses to the LLC take between 139 and 145 cycles; accesses to main memory take 148 to 158 cycles. 
These distinct distributions of access latency form the basis of cache attacks, 
as they enable an attacker to tell which part of the memory subsystem served a particular memory request. 
Our results indicate that FPGA-based attackers can precisely distinguish memory responses served by the LLC from those served by main memory.
		
\begin{figure}[tb]
			\includegraphics[width=\linewidth]{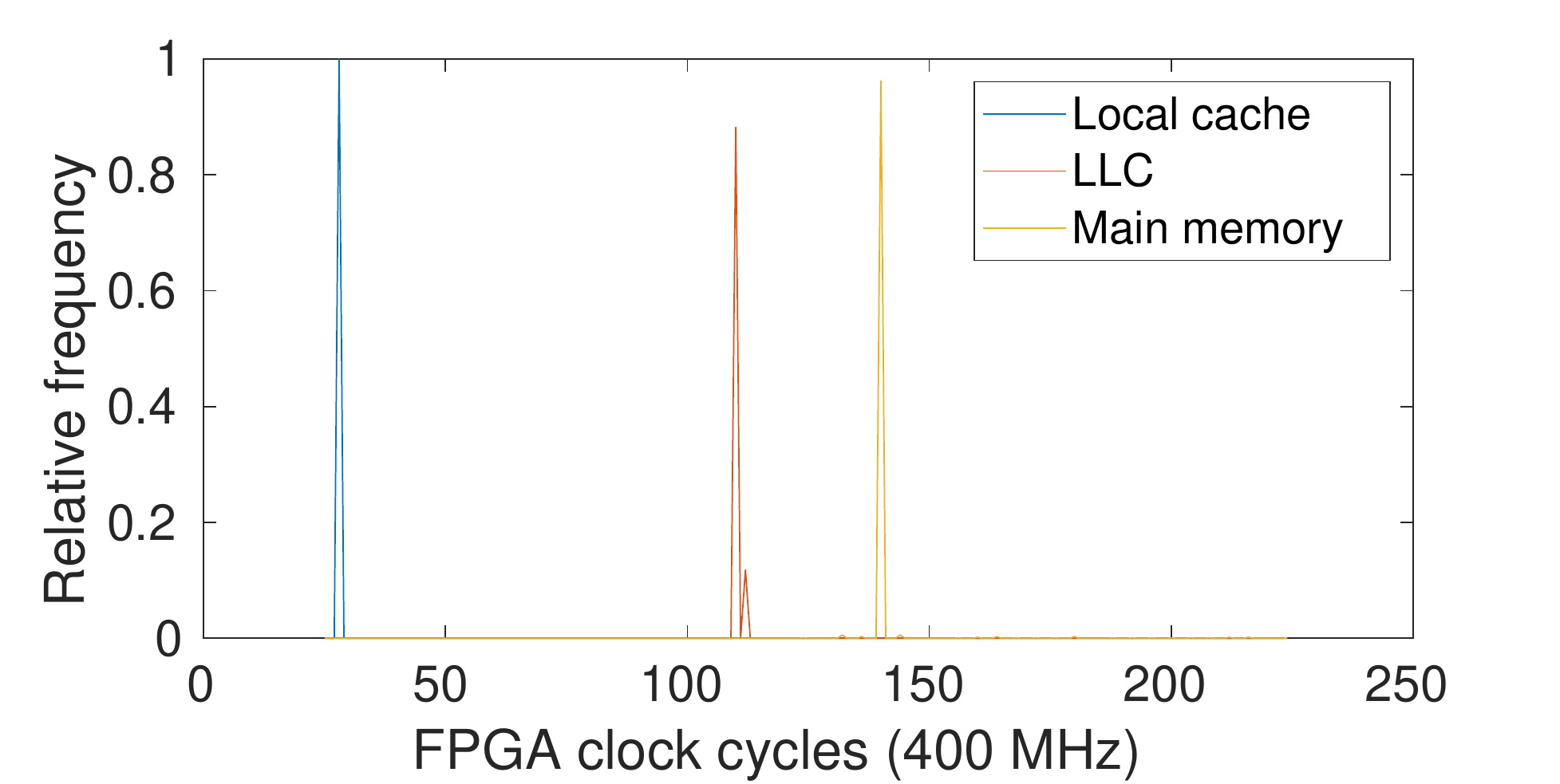}
		}{
			\caption{Latency for UPI read requests on an integrated \textsf{Arria 10} served by the FPGA's local cache, CPU's LLC, or main memory.}
			\label{fig:fpga_cache_hierarchy_timing}
\end{figure}

In addition to probing, some way of influencing the state of the cache is needed to perform cache attacks. 
We investigated all possibilities of cache interaction offered by the \textsf{CCI-P} interface on an FPGA PAC and found that 
cache lines read by the AFU from the main memory will not get cached.
While this behavior is not usable for cache attacks, it boosts Rowhammer performance as we saw in \autoref{sec:Rowhammer}.
On the other hand, cache lines written by an AFU on the PAC end up in the LLC with nearly 100\% probability.
The reason for this behavior was already discussed together with the analysis of the caching hints.
This can be used to evict other cache lines from the cache and perform eviction based attacks like Evict+Time, Evict+Reload, and Prime+Probe.
For E+T, DMA writes can be used to evict a cache line and our hardware timer measures the victim's execution time.
Even though an AFU cannot load data into the LLC, E+R can be performed as the purpose of 
reloading a cache line is to learn the latency and not literally reloading the cache line.
So the primitives for E+R on the FPGA are DMA writes and timing DMA reads with a hardware timer.
P+P can be performed using DMA writes and timing reads.
In the case where DDIO limits the number of accessible ways per cache set, other DDIO-enabled peripherals are attackable.
Flush-based attacks like Flush+Reload or Flush+Flush cannot be performed by an AFU as CCI-P does not offer a flush instruction.

\subsection{Cache Attacks from \textsf{Integrated Arria 10} FPGA to CPU}\label{sec:ca_integrated_cpu}
The integrated \textsf{Arria 10} has access to two PCIe lanes (each functioning much like the PCIe lane on the FPGA PAC) 
and one UPI lane connecting it to the CPU's memory subsystem. 
It also has its own additional cache on the FPGA accessible over UPI (cf.\ \autoref{sec:background_mem_cache_architecture}).

By timing memory requests from the AFU using our hardware timer, 
we show that distinct delays for the different levels of the memory subsystem exist. 
Both PCIe lanes have delays similar to those measured on a PAC (cf.\ \autoref{fig:pac_llc_hit_miss}). 
Our memory access latency measurements for the UPI lane, depicted in \autoref{fig:fpga_cache_hierarchy_timing}, 
show an additional peak for requests being answered by the FPGA's local cache. 
The two peaks for LLC and main memory accesses are likely narrower and further apart than in the PCIe case because UPI, Intel's proprietary high-speed processor interconnect, is an on-chip and inter-CPU bus only connecting CPUs and FPGAs. 
On all interfaces, read requests, again, are not usable for evicting cache lines from the LLC. 
DMA writes, however, can be used to alter the LLC on the CPU.
Because the UPI and PCIe lanes behave much like the PCIe lane on a PAC, 
we state the same attack scenarios (E+T, E+R, P+P) to be viable on the integrated \textsf{Arria 10}.

\subsubsection{Constructing a Covert Channel from AFU to CPU}
\begin{figure*}[tb]
  \centering
	\includegraphics[width=.8\linewidth]{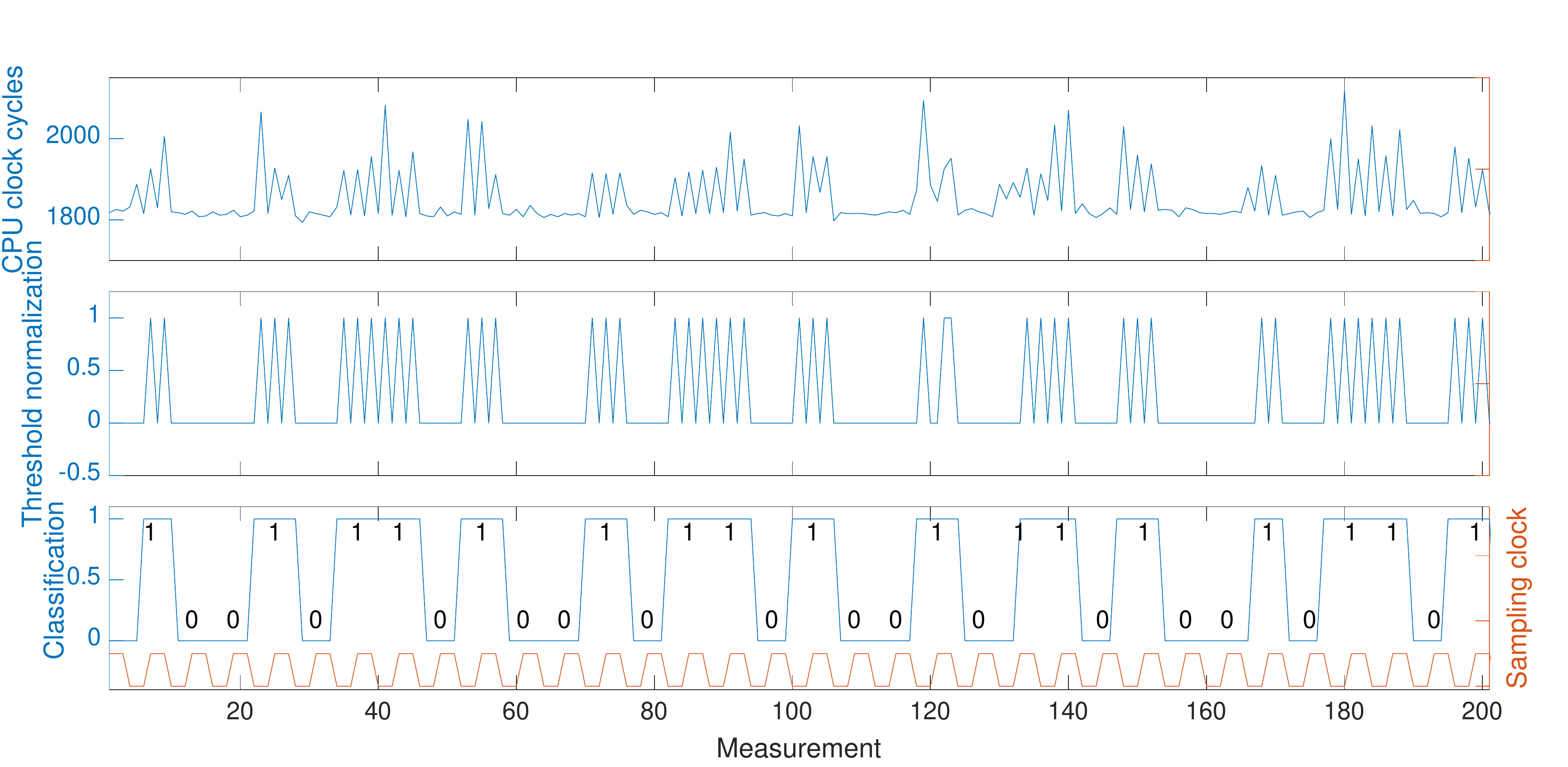}
	\caption{Covert channel measurements and decoding. The AFU sends each bit three times, which results in three peaks at the receiver if a `\texttt{1}' is transmitted (middle plot).}
	\label{fig:covert_channel_transmission}
\end{figure*}
The fact that an AFU can place data in at least one way per LLC slice allows us to construct a covert channel 
from the AFU to a co-operating process on the CPU using side effects of the LLC. 
To do so, we designed an AFU that writes a fixed string to a pre-configured cache line whenever 
a `\texttt{1}' is transmitted and stays quiet whenever a `\texttt{0}' is sent.
Using this technique, the AFU sends messages which can be read by the CPU. 
For the rest of this section, we will refer to the address the AFU writes to as the \emph{target address}.

The receiver process\footnote{This process is not the software process directly communicating with the AFU over \textsf{OPAE}/\textsf{CCI-P}.} 
first constructs an eviction set for the set/slice-pair the target address is in. 
To find an eviction set, we run a slightly modified version of Algorithm 1 using Test 1 in~\cite{vila2018theory}. 
Using the \textsf{OPAE} API to allocate hugepages and get physical addresses (cf.\ \autoref{sec:dma}) 
allows us to construct the eviction set from a rather small set of candidate addresses all belonging to the same set.

We construct the covert channel on the integrated platform as the LLC of the CPU is inclusive. 
Additionally, the receiver has access to the target address via shared memory to have the receiver 
test its eviction set against the target address directly. 
This way, we do not need to explicitly identify the target address's LLC slice. 
In a real-world scenario, either the slice selection function has to be 
known~\cite{hund2013practical,irazoqui2015systematic,inci2015seriously} or 
eviction sets for all slices have to be constructed by seeking conflicting addresses~\cite{liu2015last,oren2015spy}. 
The time penalty introduced by monitoring all cache sets can be prevented by multi-threading.

Next, the receiver primes the LLC with the eviction set found and probes the set in an endless loop. 
Whenever the execution time of a probe is above a certain threshold, the receiver assumes that the 
eviction of one of its eviction set addresses was the result of the AFU writing to the target address 
and therefore interprets this as receiving a `\texttt{1}'. 
If the probe execution time stays below the threshold, a `\texttt{0}' is detected as no eviction of the eviction set addresses occurred.
An example measurement of the receiver and its decoding steps are depicted in \autoref{fig:covert_channel_transmission}.

To ease decoding and visualization of results, the AFU sends every bit thrice and the CPU uses six probes to detect all three repetitions.
This high level of redundancy comes at the expense of speed, as we achieve a bandwidth of about \unit[94.98]{kBit/s}, 
which is low when compared to other work~\cite{wu2012whispers,liu2015last,gruss2017hello}.
The throughput can be increased by reducing the three redundant writes per bit from the AFU as 
well as by increasing the transmission frequency further to reduce the redundant CPU probes per AFU write. 
Also, multiple cache sets can be used in parallel to encode several bits at once.
The synchronization problem can be solved by using one cache set as the clock, where the AFU writes an alternating bit pattern~\cite{taram2019packetchasing}.
An average probe on the CPU takes 1855 clock cycles. 
With the CPU operating in the range of \unit[2.8 -- 3.4]{GHz}, this results in a theoretic throughput of \unit[1.5 -- 1.8]{MBit/s}.
On the other side, the AFU can on average send one write request every 10 clock cycles without filling 
the CCI-P PCIe buffer and thereby losing the write pattern.
In theory, this makes the AFU capable of sending \unit[40]{MBit/s} over the covert channel when clocked at \unit[400]{MHz}.\footnote{
This is a worst-case scenario where every transmitted bit is a `1'-bit. For a random message, this estimation goes up again as `0`-bits do not fill the buffer, allowing for faster transmission.}

Even though caching hints for memory writes are being ignored by the Blue Region, 
an AFU can place data in the LLC because the CPU is configured to handle write requests as if \texttt{WrPush\_I} is set, 
allowing for producing evictions in the LLC.
We corroborated our findings by establishing a covert channel between the AFU and the CPU with a bandwidth of \unit[94.98]{kBit/s}.
By exposing physical addresses to the user and by enabling hugepages, \textsf{OPAE} further eases eviction set finding from userspace.

\subsection{Cache Attacks from CPU to \textsf{Integrated Arria 10} FPGA}
\begin{figure*}[tb]
  \centering
  \begin{subfigure}[t]{.48\linewidth}
    \centering
 \includegraphics[width=\linewidth]{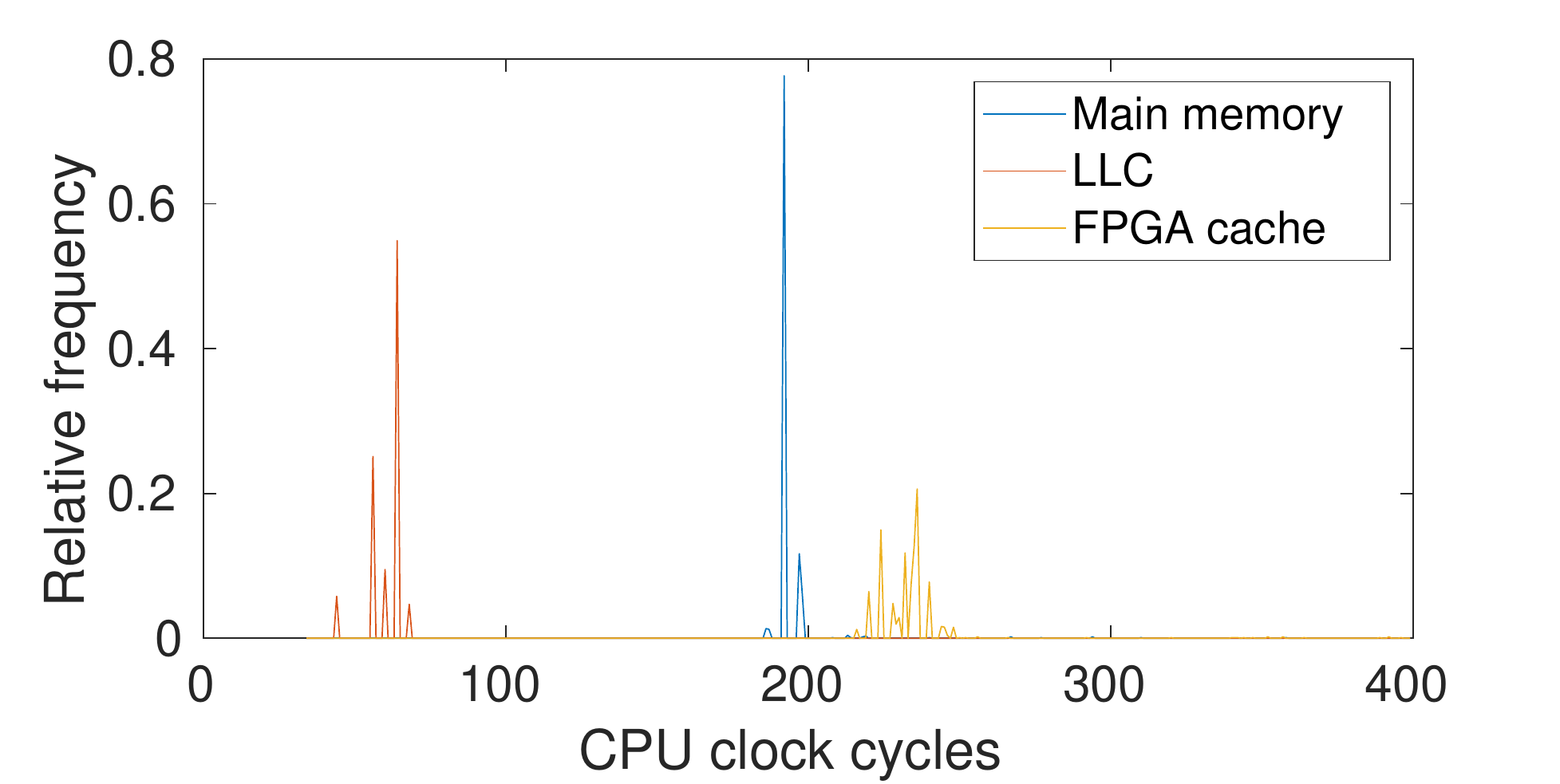}
    \caption{Memory access latency on the CPU with data being present in FPGA local cache, CPU LLC, or main memory.}
    \label{fig:cpu_cache_hierarchy_timing}
  \end{subfigure}\quad
  \begin{subfigure}[t]{.48\linewidth}
    \centering
 \includegraphics[width=\linewidth]{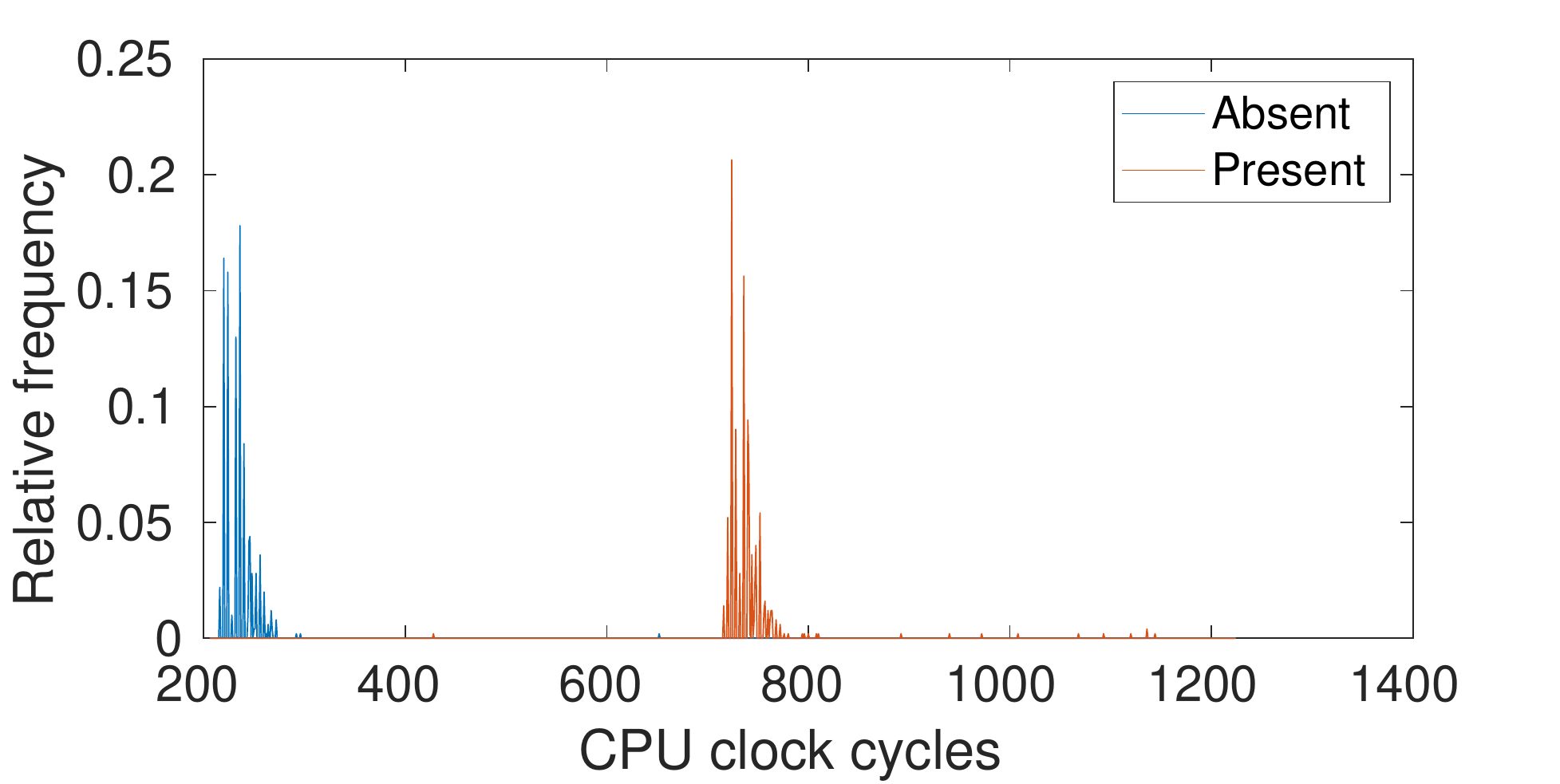}
    \caption{The \lstinline|flush| execution time on the CPU with the flushed address being absent or present in the FPGA cache.}
    \label{fig:cpu_flush_timings}
  \end{subfigure}
  \caption{Memory access and \lstinline|flush| execution latency measured from a Broadwell CPU with integrated \textsf{Arria 10}\vspace{-3ex}.}
\label{fig:cpu_latency}
\end{figure*}

We also investigated the CPU's capabilities to run cache attacks against the coherent cache on the integrated \textsf{Arria 10} FPGA. 
First, we measured the memory access latency depending on the location of the address accessed using the \texttt{rdtsc} instruction. 
The results in \autoref{fig:cpu_cache_hierarchy_timing} show that the CPU can clearly distinguish where an accessed address is located. 
Therefore, the CPU is capable of probing a memory address that may or may not be present in the local FPGA cache. 
It is interesting to note that requests to main memory return faster than those going to the FPGA cache. 
This can be explained by the much slower clock speed of the FPGA running at \unit[400]{MHz} while the CPU operates at \unit[1.2--3.4]{GHz}. 
Another explanation is that our test platform is one of the prototypes and the coherency protocol implementation of the Blue Region is still buggy. 
As nearly all known cache attack techniques rely on some form of probing phase, the capability to distinguish location of data is a good step in the direction of having a fully working cache attack from the CPU against the FPGA cache.

Besides the capability of probing the FPGA cache, we also need a way of flushing, priming, or evicting cache lines to put the FPGA cache into a known state.
While the AFU can control which data is cached locally by using caching hints, there is no such option documented for the CPU.
Therefore, priming the FPGA cache to evict cache lines is not possible.
This disables all eviction-based cache attacks.
However, as the CPU has a \texttt{clfush} instruction, we can use it to flush cache lines from the FPGA cache, because it is coherent with the LLC.
\daniel{This statements undermines the fact that you are not doing a prime+probe attack on AFU. Why can't we just prime the target llc to kick out the FPGA cache line as well?! Maybe that way, then you can run a evict+reload attack by timing the FPGA interface.}\zane{we can't prime the FPGA cache from the CPU, right? only evict?}\thore{Right, the CPU is not capable of priming the FPGA cache. But the CPU definitely can flush addresses from the FPGA cache. I'm not sure about evicting by priming the LLC right now...}
Hence, we can flush and probe cache lines located in the FPGA cache. This enables us to run a Flush+Reload attack against the victim AFU where the addresses used by the AFU get flushed before the execution of the AFU. After the execution, the attacker then probes all previously flushed addresses to learn which addresses were used during the AFU execution.
Another possible cache attack is the more efficient Flush+Flush attack. Additionally, we expect the attack to be more precise as flushing a cache line that is present in the FPGA cache takes about 500 CPU clock cycles longer than flushing a cache line that is not (cf.\ \autoref{fig:cpu_flush_timings}), while the latency difference between memory and FPGA cache accesses adds up to only about 50-70 CPU clock cycles.

In general, the applicability of F+R and F+F is limited to shared memory scenarios.
For example, two users on the same CPU might share an instantiation of a library that uses an AFU for acceleration of a process that should remain private, like training a machine learning model with confidential data or performing cryptographic operations.

\subsection{Intra-FPGA Cache Side-Channels}
As soon as FPGAs support simultaneous multi-tenancy, that is, the capability to place two AFUs from different users on the same FPGA at the same time, the possibility of intra-FPGA cache attacks arises.
As the cache on the integrated \textsf{Arria 10} is directly mapped and only \unit[128]{kB} in size, finding eviction sets becomes trivial when giving the attacker AFU access to huge pages.
As this is the default behavior of the \textsf{OPAE} driver when allocating more than one memory page at once, we assume that it is straightforward to run eviction based attacks like Evict+Time or Prime+Probe against a neighboring AFU to e.g.\ extract information about a machine learning model.
Flush-based attacks would still be impossible due to the lack of a flush instruction in CCI-P.
\daniel{How hard would it be to build to AFU to communicate over this channel? and reporr}\zane{we don't necessarily have a realistic model for multi-tenancy at the time, that's the main impediment right?}\thore{Right. The FPGA cache is directly mapped. This should make any covert channel or attack between AFUs easily implementable. But since we don't have a framework for multi-tenancy, we cannot simply test it.}

\section{Countermeasures}\label{sec:countermeasures}
\vspace{-2ex}
\para{Hardware Monitors}
Microarchitectural attacks against CPUs leave traces in hardware performance counters (HPCs) like cache hit and miss counters. 
Previous works have paired these HPCs with machine learning techniques to build real-time detectors for these attacks~\cite{briongos2018cacheshield,chiappetta2016real,zhang2016cloudradar,gulmezoglu2019fortuneteller}.
In some cases, CPU HPCs may be able to trace incoming attacks from FPGAs.
While HPCs do not exist in the same form on the \textsf{Arria 10 GX} platforms, they could be implemented by the FIM. 
A system combining FPGA and CPU HPCs could provide thorough monitoring of the FPGA-CPU interface.

\para{Increasing DRAM Row Refresh Rate}
An approach to reduce the impact of Rowhammer is increasing the DRAM refresh rate. 
DDR3 and DDR4 specifications require that each row is refreshed at least every \unit[64]{ms}, 
but many systems can be configured to refresh each row every 32 or 16 ms for better memory stability. 
When we measured the performance of our fault injection attack in \autoref{sec:rsa_fault_Rowhammer}, 
we measured the performance with varying intervals between evictions of the targeted data, 
simulating equivalent intervals in row refresh rate, since each eviction causes a subsequent 
row refresh when the memory is read by the victim program. 
\autoref{tab:performance} shows that under 1\% of attempted Rowhammers from both CPU and FPGA 
were successful with an eviction interval of \unit[32]{ms}, compared to 14\% of CPU attacks and 26\% of FPGA attacks with an interval of \unit[64]{ms}, 
suggesting that increasing the row refresh rate would significantly impede even the more powerful FPGA Rowhammer. 

\para{Cache Partitioning and Pinning}
Several cache partitioning mechanisms have been proposed to protect CPUs against cache attacks. 
While some are implementable in software~\cite{kim2012stealthmem,ye2014coloris,zhou2016software,kiriansky2018dawg} others require hardware support~\cite{liu2016catalyst,green2017autolock,gruss2017tsx}. 
When trying to protect FPGA caches against cache attacks, hardware-based approaches should be taken into special consideration. 
For example, the FIM could partition the FPGA's cache into several security domains, 
such that each AFU can only use a subset of the cache lines in the local cache. 
Another approach would introduce an additional flag to the \textsf{CCI-P} interface telling 
the local caching agent which cache lines to pin to the cache.

\para{Disabling Hugepages and Virtualizing AFU Address Space}
Intel is aware of the fact that making physical addresses available to userspace through \textsf{OPAE} 
has negative security consequences~\cite{intel2017opae}. 
Additionally to exposing physical addresses, \textsf{OPAE} makes heavy use of hugepages 
to ensure physical address continuity of buffers shared with the AFU. 
However, it is well known that disabling hugepages increases the barrier of finding eviction sets~\cite{irazoqui2015s,liu2015last} 
which in turn makes cache attacks and Rowhammer more difficult. 
We suggest disabling \textsf{OPAE}'s usage of hugepages. 
To do so, the AFU address space has to be virtualized independent of the presence of virtual environments.

\para{Protection Against Bellcore Attack}
Defenses against fault injection attacks proposed in the original Bellcore whitepaper~\cite{boneh1997importance} 
include verifying the signature before releasing it, and random padding of the message before signing, 
which ensures that no unique message is ever signed twice and that the exact plaintext cannot be easily determined.
OpenSSL protects against the Bellcore attack by verifying the signature with its plaintext and public key and 
recomputing the exponentiation by a slower but safer single exponentiation instead of by the CRT if verification does not match~\cite{carre2018openssl}. 
After we reported the vulnerability to \textsf{WolfSSL}, they issued a patch in version 4.3.0 including a  
signature verification to protect against Bellcore-style attacks.



\section{Conclusion}

In this work, we show that modern FPGA-CPU hybrid systems can be more vulnerable to well-known hardware attacks that are traditionally seen on CPU-only systems.
We show that the shared cache systems of the \textsf{Arria 10 GX} and its host CPU present possible CPU to FPGA, FPGA to CPU, and FPGA to FPGA attack vectors. 
For Rowhammer, we show that the \textsf{Arria 10 GX} is capable of causing more DRAM faults in less time than modern CPUs.
Our research indicates that defense against hardware side-channels is just as essential for modern FPGA systems as it is for modern CPUs. 
Of course, the security of any device physically installed in a system, like a network card or graphics card, is important, but 
FPGAs present additional security challenges due to their inherently flexible nature. 
From a security perspective, a user-configurable FPGA on a cloud system needs to be treated with at least as much care and caution as a user-controlled CPU thread, as it can exploit many of the same vulnerabilities.

\ifdefined\PUBLICATION
\para{Acknowledgments}
This research received partial funding, hardware donations, and extremely useful advice from Intel and its employees. 
We would like to especially thank Alpa Trivedi from Intel.
Daniel Moghimi was supported by the National Science Foundation under grant no. CNS-1814406. This work was partially supported by the German Research Foundation (DFG) project number 427774779.
\fi

{\footnotesize \bibliographystyle{plain}
  \sloppy
\bibliography{references}}

\end{document}